\documentclass[final,authoryear,2p,12pt]{elsarticle}
\usepackage{geometry}                
\usepackage{graphicx}
\usepackage{epstopdf}
\usepackage{booktabs}
\usepackage{natbib}

\usepackage{hyperref}
\usepackage{amsthm}
\usepackage{cleveref}

\usepackage{amsmath}
\usepackage{amsfonts}
\usepackage{multirow}
\usepackage{lscape}
\usepackage{bbm}
\usepackage{bm}
\usepackage{color}

\usepackage{amssymb}
\usepackage{xcolor}

\usepackage{color,soul}
\definecolor{lightblue}{rgb}{.80,.95,1}
\sethlcolor{lightblue}
\setulcolor{red}

\begin{document}
\begin{frontmatter}

\title{Co-jumping of Treasury Yield Curve Rates}

\author[ies,utia]{Jozef Barun\'ik\corref{cor2}}
\author[ies,utia]{Pavel Fi\v{s}er}

\address[ies]{Institute of Economic Studies, Charles University in Prague, Opletalova 26, 110 00 Prague, Czech Republic}
\address[utia]{The Czech Academy of Sciences, Institute of Information Theory and Automation, Pod Vodarenskou Vezi 4, 182 00 Prague, Czech Republic}
\cortext[cor2]{Corresponding author, Tel. +420(776)259273, Email address: barunik@fsv.cuni.cz}
\tnotetext[label1]{We are indebted to Torben Andersen, Ionut Florescu, Giampiero Gallo, Roberto Ren\`{o}, Lukas Vacha, David Veredas, and participants at various seminars and conferences for many useful comments, suggestions and discussions. Support from the Czech Science Foundation under the 19-28231X (EXPRO) project is gratefully acknowledged. For estimation of the co-jumps as well as other quantities introduced by this paper, we provide the code \texttt{waveletcojumps} in \textsf{R} software. The code is available on \url{https://github.com/FiserPavel/waveletcojumps}}

\begin{abstract}
We study the role of co-jumps in the interest rate futures markets. To disentangle continuous part of quadratic covariation from co-jumps, we localize the co-jumps precisely through wavelet coefficients and identify statistically significant ones. Using high frequency data about U.S. and European yield curves we quantify the effect of co-jumps on their correlation structure. Empirical findings reveal much stronger co-jumping behavior of the U.S. yield curves in comparison to the European one. Further, we connect co-jumping behavior to the monetary policy announcements, and study effect of 103 FOMC and 119 ECB announcements on the identified co-jumps during the period from January 2007 to December 2017.
\end{abstract}
\end{frontmatter}

\section{Introduction}

The understanding of the financial market behavior is fundamental not only for investors but also for policy makers and researchers alike, and it lies at heart of financial research for decades. Behavior of bond prices is central to this research since bond market reflects monetary policy decisions about short term interest rates taken by central banks. This information is then transferred to the long end of the yield curve which in turn has influence on the future development of the whole economy. Yield spreads are used to forecast future short yields, real activity or inflation, and these forecasts further form a basis information set used for investment, saving and policy decisions. Moreover, the bond prices play essential role in valuing number of financial derivatives as swaps, futures, options and construction of hedging strategies. A very important question to ask is then how sensitive the bond markets are to exogenous events creating jumps, and co-jumps along the treasury yield curve rates. With a few exceptions discussed below, much of recent work has approached bond markets assuming no significant co-jump behavior silently.

In this paper, we aim to identify the role of co-jumps in the U.S., and European Treasury markets, and contribute to the empirical literature studying bond markets in several ways. First, we aim to find how co-jumps impact correlation structure of the treasury markets. Second, using high frequency data about the interest rate futures and precise localization techniques, we would like to find what portion of co-jumps can be identified by macroeconomic announcements in both markets. Third, we would like to document response patterns of the whole yield curve to monetary policy announcements. Finally, using equal sample period of the high frequency data about the U.S. and European interest rate markets, we would like to study the differences in the co-jumps behavior, their role, and reactions at the two important bond markets. 

The importance of jumps is studied vastly in the univariate setting using single maturities. \citet{johannes_2004} finds that jumps play an essential role in describing interest rate dynamics as they account for a substantial part of total volatility. Moreover, \citet{andersen_2007,busch_2011,sheppard_2015} show that incorporating the jump component improves out-of-sample volatility forecast. Besides the benefits for volatility modeling, vast literature looks at role of jumps in connection to macroeconomic news announcements. From many, \citet{Balduzzi_1997}, \citet{Balduzzi_2001}, \citet{das_2002}, \citet{piazzesi_2005} consistently report that Federal Open Market Committee (FOMC) meetings and scheduled macroeconomic news announcements are an important factor in explaining jumps.

While it is important to understand the role of jumps in interest rate markets, it is tempting to find  effects of co-jumps on the whole term structure of the yield curve as co-jumps are able to capture the co-movements of prices across term structure. This distinction provides new insights about bond market dynamics as the detection of co-jumps across different maturities provides additional information compared to the case where individual bond maturities jump in isolation. Yet, literature is limited with this respect. Among few, \citet{CJ_ECB_2014} examine structure of the yield curve in reaction to announcements of ECB using two distant terms of German government bond futures from 2001 to 2012. \citet{J_Dungey_2009} study the same problem from the perspective of US Treasury market and scheduled FED announcement on data from 2002 to 2006. \citet{CJ_Dungey_2012} distinguish from previous literature by examining co-jumping between spot and futures in the US Treasury market. Moreover, \citet{lahaye_2011} relate the co-jumps extracted from a panel of US stocks to US macroeconomic announcements and model how news surprises explain co-jumps.

Whereas these contributions provide first results on behavior of co-jumps, the role of co-jumps in the treasury yield curve is not fully understood. One of the important question is how jumps and co-jumps influence correlation structure of the yield curve. Literature has studied role of co-jumps in asset \citep{clemens_2013} or currency markets \citep{BV_2017} reporting significant positive impact on correlations, but with few exceptions mentioned above, literature stay silent about bond markets, especially in connection to high frequency data. Since mentioned papers use very distinct methodologies as well as time spans, the results are also not comparable directly. Moreover, methods are often not fully effective in separation of co-jumps and large disjoint jumps.

In this study, we contribute to the growing literature by estimating co-jumps using precise techniques, and comprehensively studying the role of co-jumps in the U.S. and European treasury yield curve rates. We rely on the recently introduced approach to identify co-jumps \citep{BV_2017} based on a wavelet decomposition of stochastic processes. The main reason why we focus on wavelet analysis is its remarkable ability to detect jumps and sharp cusps even if covered by noise \citep{donoho_1994,wang95}. The reported improvements originate from the fact that wavelets are able to decompose noisy time series into separate time-scale components. This decomposition then helps to distinguish jumps from continuous price changes, and microstructure noise effects as wavelet coefficients decay at a different rate for continuous and jump processes. Wavelet coefficients at jump locations are larger in comparison to other observations. While changes in continuous price processes over a given small time interval are close to zero, changes in jumps are significantly larger. Wavelet coefficients are able to precisely distinguish between these situations, and hence locate jumps very precisely. Moreover, the method is able to further decompose the discontinuous process and separate jumps and co-jumps while minimizing the identification of false co-jumps resulting from presence of large disjoint jumps. 

For the empirical analysis, we use the tick data about interest rate futures. There are multiple reasons why to analyze futures instead of cash market in order to examine co-jump behavior of the yield curve. Importantly, availability of high frequency tick data for the futures contracts makes futures advantageous in comparison to cash market. Futures markets are becoming more liquid over time, and are gaining importance relative to cash market. Due to delivery mechanism of the U.S. Treasury futures contracts, futures prices are tightly linked to underlying bond prices (and yields), and moreover, also due to lower transaction costs, futures market was detected to be dominant to cash market in reaction to news and price discovery process \citep{brandt2007price}.

Empirical results provide interesting evidence about role of the co-jumps in the interest rate markets. In a nutshell, collecting tick by tick data about the U.S. and European interest rate futures market for the period spanning the dates 2007 -- 2017, and using precise wavelet-based estimators to identify co-jumps, we document approximately twice larger role of co-jumps in the U.S. market in comparison to European market. The role of co-jumps changes dynamically over time, and we find co-jumps to significantly inflate the correlation structure. We also document high probability of co-jump occurrence during the announcement days using the data about 103 FOMC and 119 ECB announcement days with the U.S. market being more responsive to the news. Finally, we disentangle two distinct response patterns, rotations and level shifts, of the yield curve to monetary policy announcements. Rotations represents the situation in which the short and long end jump in the opposite directions and can be interpreted as change of central bank policy preferences. Level shifts, on the other hand, can be described as a situation in which market interprets the news announcement as a policy surprise and as a consequence all yields across maturities jump in the same direction. We document that approximately 30\% of the news are accompanied by level shift of the whole yield curve in both markets. On the contrary, only 1\% of announcements are interpreted by the market as an adjustment in preferences.

\section{Data}
\label{data_description}

The main aim of the paper is to study the jump and co-jump behavior of interest rate futures prices on the U.S. and European interest rate markets, as well as impact of news announcements on the co-jumps. Here, we describe the data about the contracts as well as news announcements data used in the analysis for the two separate markets.

\subsection{The U.S. Treasury bond futures contracts}

The U.S. Treasury bond futures contracts are traded on the Chicago Mercantile Exchange (CME) and comprise tick by tick data for transaction prices and volume. We examine (active) contracts for each benchmark tenors, i.e. 2-year, 5-year, 10-year, and 30-year bond maturity with symbols TU, FV, TY, and US. The data are available from Tick Data, Inc.\footnote{\href{http://www.tickdata.com/}{{http://www.tickdata.com/}}} Since December 18, 2006 CME offers almost continuous trading using a Globex\textregistered  electronic platform with 23 hours trading day from Sunday until Friday. The trading hours start at 17:00 of the previous day and end at 16:00 U.S. Central Standard Time (CST) when the trading is interrupted with a sixty minute break. Since introduction of the trading system dramatically affected trading activity, we restrict our sample period to trading days extending from January 5, 2007 through December 28, 2017. 

\begin{figure}[!ht]
\centering
\includegraphics[width=0.8\textwidth]{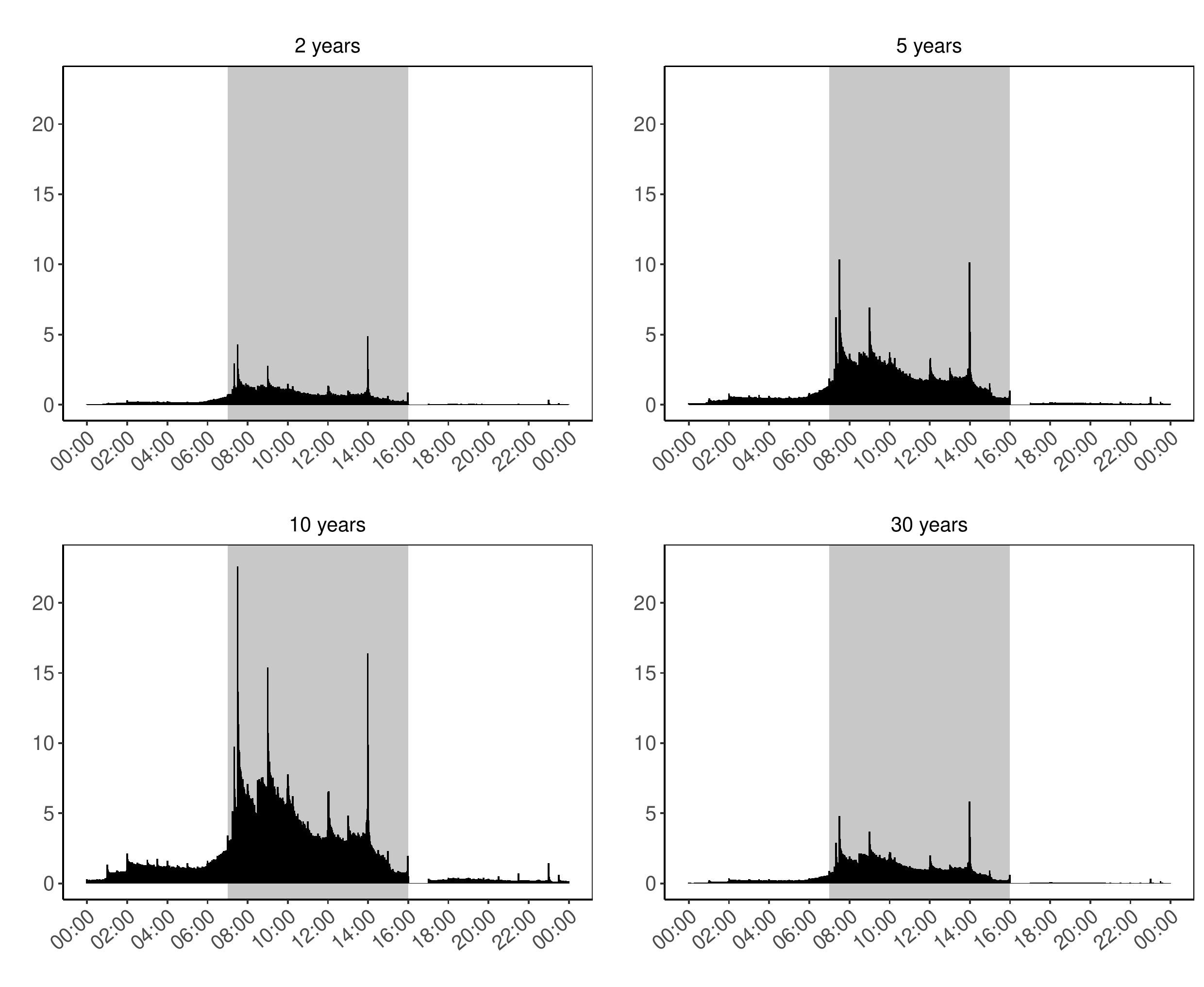}
\caption{Trading activity on 2 year, 5 year, 10 year, and 30 year US bond futures contracts measured in terms of the total volume (scaled by $10^6$ in the figure) using 1-min trading intervals over the whole period of January 5, 2007 to December 28, 2017. Trading session used for the analysis is shaded. }  
\label{Fig_volume}    
\end{figure}

It is interesting to look at the trading activity of the futures during different times of the day. Trading activity is measured on one minute intervals as a total traded volume over the whole sample. Figure ~\ref{Fig_volume} shows low trading activity after CME opening hours representing the Asia session (17:00 - 2:00). With the start of European session (2:00 - 8:00) the measured volume slightly increases and peaks when both European and US market are actively trading. The peak is followed by the US session with the volumes on the highest levels. From the perspective of bond futures, by far the 10 year maturity bonds displays the largest trading activity, followed by the 5 year bonds and the 30 year bonds respectively. Oppositely the lowest volumes are identified for 2 year bonds which can be caused by the period of Quantitative easing where the Federal Fund rates were at the zero lower bound.  

In order to get unbiased estimates we exclude illiquid part of the 23 hour trading hours from the analysis. We define a trading day lasting for 9 hours (7:00 - 16:00) where 83\% of all trade volume is executed and covers the main U.S. trading hours with one hour overlap to the European session. 

Moreover, weekends, U.S. federal holidays, periods between December 24 - December 26 and December 31 - January 2 and days identified as low trading days\footnote{Days during which trade occurs in less then 60 percent of 5 minute observations consistently across all four US bond futures time series.} 
are removed from the dataset. The final dataset, after excluding non-trading and low-trading days, consists of 2705 trading days. Finally, the data are synchronized by sampling the futures prices at one minute and five minute intervals taking the last available transaction within the interval as an observation for this time stamp. 

Table \ref{tab_descriptive_stat} captures the descriptive statistics of logarithmic futures returns with one minute and five minute frequencies used in the analysis. The average returns increase with increasing maturity and with lower sampling frequency i.e. the largest mean returns are obtained for bond futures with 30 year maturity with 5 minute sampling frequency. The same applies for standard deviation - with increasing maturity and lower sampling frequency standard deviation increases.

\begin{table}
\footnotesize
\begin{tabular}{@{}lcrrrrcrrrrc@{}}\toprule
  & \phantom{a} & \multicolumn{4}{c}{1 min} &\phantom{ab} & \multicolumn{4}{c}{5 min} & \phantom{a} \\  
  \cmidrule{3-6} \cmidrule{8-11} 
  & & 2y & 5y & 10y & 30y & & 2y & 5y & 10y & 30y & \\ 
  \midrule
  Mean & & 0.088 & 0.183 & 0.234 & 0.235 &  & 0.442 & 0.914 & 1.169 & 1.174 &\\ 
  Min & & -0.003 & -0.006 & -0.010 & -0.013 &  & -0.003 & -0.005 & -0.009 & -0.014 &\\ 
  Max & & 0.003 & 0.017 & 0.031 & 0.048 &  & 0.003 & 0.011 & 0.022 & 0.034& \\ 
  Std.Dev & & 0.477 & 1.007 & 1.613 & 2.756 &  & 0.868 & 2.092 & 3.254 & 5.619 &\\ 
  Skewness & & -0.158 & 2.598 & 4.534 & 3.140 &  & -0.029 & 0.604 & 1.046 & 0.563 &\\ 
  Kurtosis & & 73.677 & 646.118 & 1034.078 & 676.523 &  & 58.108 & 73.181 & 101.843 & 66.028 &\\ 
   \bottomrule
\end{tabular}
\caption{Descriptive statistics for the U.S. bond futures logarithmic price returns with 2, 5, 10, 30 year maturity. Descriptive statistics are shown for 1 minute and 5 minute frequency of intraday returns (7:30 - 16:00) for the whole sample period - from January 5, 2007 to December 28, 2017. Means are scaled by $10^5$, and standard deviations are scaled by $10^3$.}
\label{tab_descriptive_stat}
\end{table}

\subsection{FOMC news announcements}

To study the impact of news on co-jumps on the interest rate futures market, we consider scheduled macroeconomic announcements - The Federal Open Market Committee (FOMC) press releases. The FOMC represents the monetary policy decision making body of the Federal Reserve system adjusting the levels of short-term interest rates in order to promote maximum employment, stable prices and moderate long-term interest rates in the U.S. economy. At the end of 2007 - 2009 Financial crisis and subsequent period, the FOMC also used the so-called ``Quantitative Easing'' based on large-scale purchase of securities issued by government in order to lower longer-term interest rates and consequently improve economic performance \citep{FED_pub_2016}. 

The Committee meets regularly eight times a year but under special circumstances may also hold unscheduled meetings as witnessed several times during the recent Financial crisis. The purpose of these meetings is to review current and future economic and financial developments. Importantly for the analysis, FOMC meetings are held for two days but the FOMC public press conference takes place only during the second day at 13:00 CST when the FOMC statement is also published. 

The dates of FOMC meetings were obtained from the Board of Governors of the Federal Reserve.\footnote{\href{https://www.federalreserve.gov/monetarypolicy/fomccalendars.htm}{https://www.federalreserve.gov}} We found 106 scheduled or unscheduled meetings during the sample period January 5, 2007 - December 28, 2017 out of them 3 dates correspond to the weekend or non-trading day and therefore were excluded. This leaves 103 FOMC announcement days used for the analysis.




\subsection{The European bond futures contracts}
\label{sec:eudata}

The European bond futures contracts are traded on the Eurex exchange in Europe - Berlin and comprise tick by tick data for transaction prices and volume. We examine (active) contracts for each benchmark tenors, i.e. 2-year (EURO-Schatz), 5-year (EURO-Bobl), 10-year (EURO-Bund), and 30-year (EURO-Buxl) bond maturity with symbols BZ, BL, BN, and BX. The data are available from Tick Data, Inc.\footnote{\href{http://www.tickdata.com/}{{http://www.tickdata.com/}}}. The sample period we use is restricted to trading days extending from January 5, 2007 through December 28, 2017 due to the same reasons mentioned above.

\begin{figure}[!ht]
\centering
\includegraphics[width=0.8\textwidth]{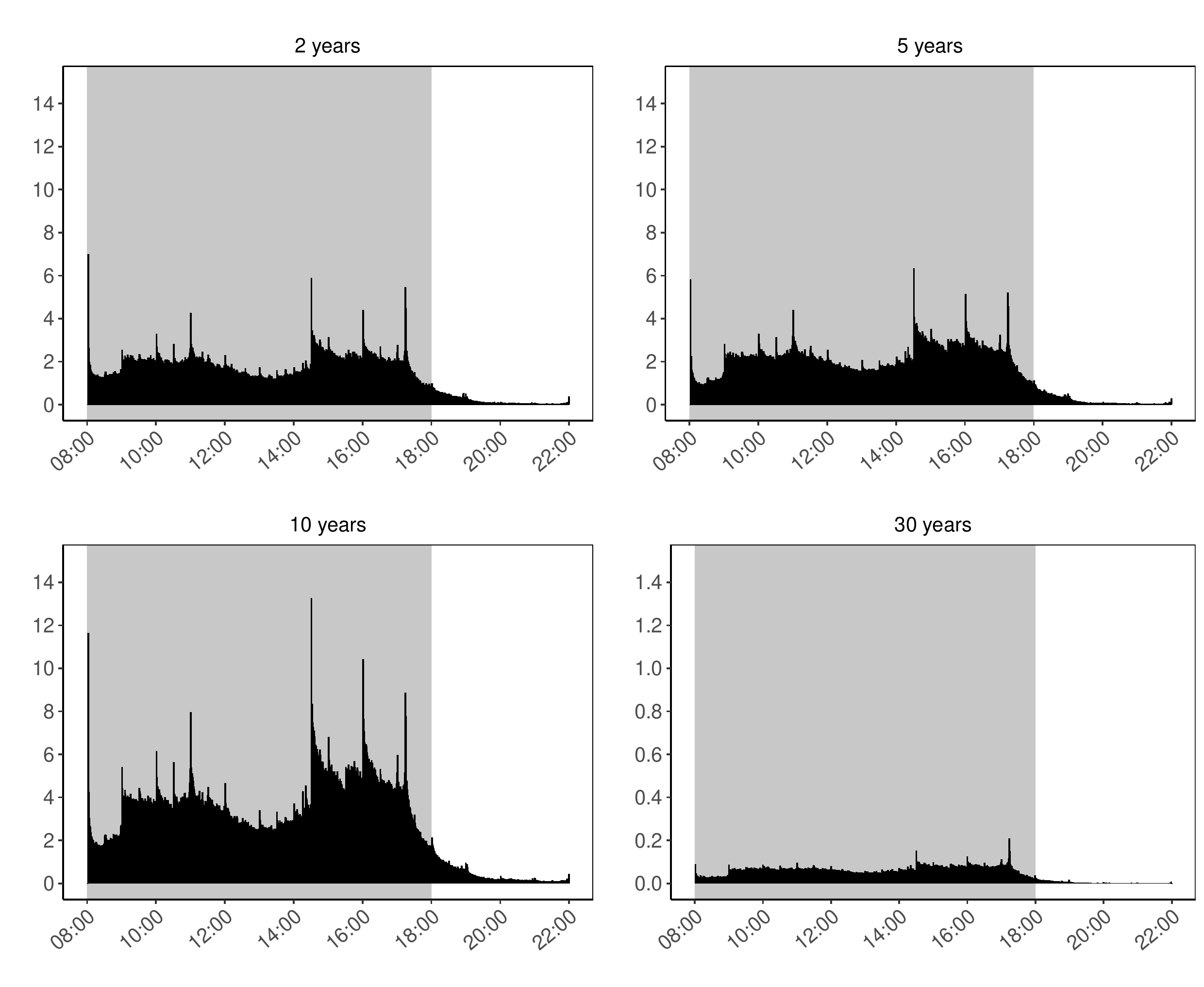}
\caption{Trading activity on 2 year, 5 year, 10 year, and 30 year European bond futures contracts measured in terms of the total volume (scaled by $10^6$ in the figure) using 1-min trading intervals over the whole period of January 5, 2007 to December 28, 2017. Trading session used for the analysis is shaded. }      
\label{Fig_volume_EU}
\end{figure}

Trading activity measured on one minute intervals as a total traded volume over the whole sample is shown in Figure \ref{Fig_volume_EU}. From the perspective of bond futures, by far the 10 year maturity bonds displays the largest trading activity, followed by the 5 year bonds and the 2 year bonds respectively. Oppositely the lowest volumes are identified for 30 year bonds. In order to get unbiased estimates we exclude illiquid part of the trading hours from the analysis, and we define a trading day lasting for 10 hours (8:00 - 18:00) of the European time where most of all trade volume is executed and covers the main trading hours. 

Similarly to the U.S. data, we remove bank holidays, periods between December 24 - December 26 and December 31 - January 2. The final dataset consists of 2780 trading days. Finally, the data are synchronized by sampling the futures prices at one minute and five minute intervals taking the last available transaction within the interval as an observation for this time stamp.

\begin{table}
\footnotesize
\begin{tabular}{@{}lcrrrrcrrrrc@{}}\toprule
  & \phantom{a} & \multicolumn{4}{c}{1 min} &\phantom{ab} & \multicolumn{4}{c}{5 min} & \phantom{a} \\  
  \cmidrule{3-6} \cmidrule{8-11} 
  & & 2y & 5y & 10y & 30y & & 2y & 5y & 10y & 30y & \\ 
  \midrule

  Mean & & 0.040 & 0.180 & 0.309 & 0.267 &  & 0.216 & 0.718 & 1.089 & 0.970 \\ 
  Min & & -0.002 & -0.004 & -0.015 & -0.015 &  & -0.002 & -0.004 & -0.015 & -0.019 \\ 
  Max & & 0.002 & 0.004 & 0.014 & 0.014 &  & 0.003 & 0.005 & 0.014 & 0.018 \\ 
  Std.Dev & & 0.392 & 0.908 & 1.444 & 3.209 &  & 0.655 & 1.630 & 2.751 & 6.199 \\ 
  Skewness & & -0.269 & -0.119 & -0.290 & -0.198 &  & -0.141 & 0.002 & -0.191 & -0.112 \\ 
  Kurtosis & & 40.433 & 27.875 & 140.464 & 43.346 &  & 42.369 & 23.929 & 46.302 & 24.589 \\
   \bottomrule
\end{tabular}
\caption{Descriptive statistics for the European bond futures logarithmic price returns with 2, 5, 10, 30 year maturity. Descriptive statistics are shown for 1 minute and 5 minute frequency of intraday returns (7:30 - 16:00) for the whole sample period - from January 5, 2007 to December 28, 2017. Means are scaled by $10^5$, and standard deviations are scaled by $10^3$.}
\label{tab_descriptive_stat_EU}
\end{table}

Table \ref{tab_descriptive_stat_EU} captures the descriptive statistics of logarithmic futures returns with one minute and five minute frequencies used in the analysis. Similarly to the U.S. interest rate market, average returns together with standard deviation increase with increasing maturity and with lower sampling frequency.

\subsection{The ECB news announcements}

Since we aim to study impact of news on co-jumps on the interest rate futures market, we consider scheduled macroeconomic announcements - The European Central Bank (ECB) press releases.

The decision on the key rate is published via a press release at 13:45 CET and is further explained in a subsequent press conference starting at 14:30 CET \cite{CJ_ECB_2014}. We take a 15 minute window after press release (13:30-13:45) and 30 minute window after 14:30 (14:30 - 15:00). Since the announcement days no more consist only from Thursdays, we take all other days as the non-announcement days. The dates of ECB announcement were obtained from the European Central Bank.\footnote{\href{https://www.ecb.europa.eu}{https://www.ecb.europa.eu}} Our data contain 119 ECB announcement days used for the analysis

\section{Estimation of co-jumps in the interest rate futures}
\label{section_estimation_QV}

We start the analysis with decomposition of interest rate futures into several components including jumps and co-jumps. The methodology we use here closely follows our previous work, where such decomposition has been proposed based on wavelet based estimation of quadratic covariation \citep{BV_2017}.

To set out the notation, consider the observed $d-$variate interest rate futures (log) price process $(\mathbf{Y}_t)_{t\in[0,T]}$ describing the $\ell=1,\ldots,d$ interest rate futures prices with different maturities $\mathbf{Y}_t=\left(Y_{t,\ell_1},\ldots,Y_{t,\ell_d}\right)'$. The common assumption about the observed prices is that we can decompose it into an underlying (log) price process $(\mathbf{X}_t)_{t\in[0,T]}$ and a zero mean, finite variance i.i.d. noise term $(\boldsymbol \epsilon_t)_{t\in[0,T]}$ capturing microstructure noise. 

The joint dynamics of the $\ell_1, \ell_2 \in \{1, \ldots, d\}$ components of the latent price process \textbf{X}$_t$ can be described by 
\begin{eqnarray}
\mathrm{d} X_{t,\ell_1}&=&\mu_{t,\ell_1}\mathrm{d}t+\sigma_{t,\ell_1}\mathrm{d}B_{t,\ell_1} + \mathrm{d} J_{t,\ell_1} \\
\mathrm{d} X_{t,\ell_2}&=&\mu_{t,\ell_2}\mathrm{d}t+\sigma_{t,\ell_2}\mathrm{d}B_{t,\ell_2} + \mathrm{d} J_{t,\ell_2},
\end{eqnarray}
where $\mu_{t,\ell_i}$ and $\sigma_{t,\ell_i}$ are c\`{a}dl\`{a}c stochastic processes, $B_{t,\ell_i}$ is a standard Brownian motion correlated with $\rho^{\ell_1,\ell_2} = corr(B_{t,\ell_1},B_{t,\ell_2})$ , and $J_{t,\ell_i}$ describes a right continuous pure jump process for $i = {1,2}$. The jump process is assumed to have finite number of jumps in a finite time interval and can be correlated. 

The quadratic return covariation of the latent interest rate futures prices $X_{t,\ell_1}$ and $X_{t,\ell_2}$ over the fixed period of time [0,T] can be decomposed into two parts - integrated covariance $IC_{\ell_1,\ell_2}$ (continuos part) and co-jump variation $CJ_{\ell_1,\ell_2}$ (discontinuous part)
\begin{equation}
QV_{\ell_1,\ell_2} = \underbrace{\int_0^T \sigma_{t,\ell_1}\sigma_{t,\ell_1}\mathrm{d}\langle B_{\ell_1},B_{\ell_2}\rangle_t}_{IC_{\ell_1,\ell_2}} + \underbrace{\sum_{0\le t \le T} \Delta J_{t,\ell_1} \Delta J_{t,\ell_2}}_{CJ_{\ell_1,\ell_2}}.
\end{equation}
Note that the co-jump variation term $CJ_{\ell_1,\ell_2}$ is non-zero only if both $\Delta J_{t,\ell_1}$ and $\Delta J_{t,\ell_2}$ are non-zero. The full quadratic covariance matrix of the interest rate futures prices $\bm{QV}$ holds elements of $\bm{IC}$ and $\bm{CJ}$ as
\begin{equation}
\label{eq_QV}
\bm{QV} = \bm{IC}+\bm{CJ}=\left(
\begin{array}{cc}
IC_{\ell_1,\ell_1} + CJ_{\ell_1,\ell_1} & IC_{\ell_1,\ell_2}+ CJ_{\ell_1,\ell_2}\\
IC_{\ell_2,\ell_1} + CJ_{\ell_2,\ell_1}& IC_{\ell_2,\ell_2}+ CJ_{\ell_2,\ell_2}
\end{array}
\right),
\end{equation}
where on the diagonal of the covariance matrix is quadratic variation for $\ell_1 = \ell_2$ and the quadratic covariation for $\ell_1 \neq \ell_2$ elsewhere. To study the co-jumps of interest rate futures, we are interested to estimate both components of the \autoref{eq_QV} - continuous covariation and co-jumps. 

A first step to estimate the quadratic covariation is to consider realized covariance that can be computed over a fixed time interval [0 $\leq $ t $\leq$ T] as 
\begin{eqnarray}
\label{eq_QV_RC}
\widehat{QV}_{\ell_1,\ell_1}^{(RC)} = \sum_{i=1}^N \Delta_i Y_{t,\ell_1} \Delta_i Y_{t,\ell_2}
\end{eqnarray}
where $\Delta_i Y_{t,\ell} = Y_{t+i/N,\ell} - Y_{t+i-1/N,\ell}$ is the i-th intraday return over the time interval [0, T]. The realized covariance estimator consistently estimates the quadratic covariation provided that the processes are not contaminated with  microstructure noise i.e. covariation associated with ($Y_{t,\ell_1}$, $Y_{t,\ell_2}$) \citep{Andersen_2003, barndorff_2004}. Since we are interested in covariance associated with ($X_{t,\ell_1}$, $X_{t,\ell_2}$), we need to apply estimators capable of recovering the covariation of the latent process. For this, \citet{BV_2017} propose to combine the two-scale covariance estimation using subsampling of the returns \citep{zhang2011} with wavelet decomposition to recover co-jumps.

To localize the co-jumps in the interest rate futures prices precisely, we use the frequency domain tools. Assuming the finite number of jumps in the process, and building on the results of \citet{fan_wang_2007} about the ability of wavelets to precisely detect jumps in deterministic functions, \citet{BV_2017} generalize this result to a continuous time setting. They first decompose the process to frequencies, and then use the highest frequency to detect sharp discontinuities in the process without being influenced by other components. Hence, the estimator of \citet{BV_2017} is able to distinguish between continuous and discontinuous parts of the stochastic price process precisely.

The co-jump variation associated with $(X_{t,\ell_1}$, $X_{t,\ell_2})$ over $[0 \leq t \leq T]$ is estimated in the discrete synchronized time as sum of co-jumps
\begin{eqnarray}
\label{eq_CJ_definition}
\widehat{CJ}_{\ell_1,\ell_2} = \sum_{i=1}^N \Delta_i J_{t,\ell_1} \Delta_i J_{t,\ell_2},
\end{eqnarray}
with jump size at intraday position \textit{i} $\Delta_i J_{t,\ell}$ being estimated as
\begin{eqnarray}
\label{eq_jump_size}
\Delta_i J_{t,\ell} =  (\Delta_i Y_{t,\ell}) \mathbbm{1}_{\vert \mathcal{W}_{1,k}^\ell \vert > \xi},
\end{eqnarray}
where $\mathcal{W}_{1,k}^\ell$ represents the intraday wavelet coefficient at the first scale\footnote{Since we estimate the quadratic covariation on discrete data, we use a non-subsampled version of a discrete wavelet transform, more specifically, the maximal overlap discrete wavelet transform (MODWT). A brief introduction of the discrete wavelet transform and MODWT can be found in \ref{dwt}.} and $\xi$ is the jump threshold. As a threshold $\xi$ we use the universal threshold introduced by \citet{donoho_1994} and used by \citet{BV_2017} with the intraday median absolute deviation estimator of standard deviation adapted for the MODWT coefficients\footnote{For details, see \cite{PercivalWalden2000}. As we use the MODWT filters, we must slightly correct the position of the wavelet coefficients to obtain the precise jump position; see \cite{PercivalMofjeld1997}.} in the following form:
\begin{eqnarray}
\label{eq_J_threshold}
\xi = \sqrt{2} \, \textrm{median} \lbrace \vert \mathcal{W}_{1,k}^\ell \vert \rbrace \sqrt{2logN} / 0.6745 .
\end{eqnarray}
If the absolute value of an intraday wavelet coefficient exceeds the threshold $\xi$, then the jump will be estimated at position $k$. In other words, the noise and the continuous part are relatively small, and hence, the dominance of $\mathcal{W}_{1,k}^{\ell}$ results from a discontinuous jump. Then, a co-jump occurs only if both jumps in process $(X_{t,\ell_1},X_{t,\ell_2})$ occur simultaneously. Since the jumps and co-jumps are estimated consistently \citep{fan_wang_2007,BV_2017}, the co-jump variation matrix associated with ($X_{t,\ell_1}$, $X_{t,\ell_2}$) has following form: 
\begin{eqnarray}
\widehat{\textbf{CJ}} =
\begin{pmatrix}
\widehat{CJ}_{\ell_1,\ell_1} & \widehat{CJ}_{\ell_1,\ell_2} \\
\widehat{CJ}_{\ell_2,\ell_1} & \widehat{CJ}_{\ell_2,\ell_2}
\end{pmatrix}.
\end{eqnarray}

\subsection{Jump wavelet covariance estimator}

Using the time synchronized, jump--adjusted interest rate futures price process $Y_{t,\ell}^{(J)}=Y_{t,\ell}-\widehat{CJ}_{\ell,\ell}$, \citet{BV_2017} propose the estimator of continuous part of quadratic covariation that is robust against noise, and hence obtain full decomposition of the covariation matrix from the observed data. In addition, one can use the estimator of \citet{BV_2017} to further decompose the covariation into time scales and study behavior of the quantities at different investment horizons. Since we are interested solely in co-jumps, we leave the discussion of the details to an interested reader.

Jump wavelet covariance estimator (JWC) of the integrated covariance of the bond futures return process $(X_{t,\ell_1}$, $X_{t,\ell_2}) \in L^2(\mathbb{R})$ over the fixed time interval $[0 \leq t \leq T]$ can be estimated from the time-synchronized jump-adjusted price process $(Y_{t,\ell_1}^{(J)}$, $Y_{t,\ell_2}^{(J)})$ as
\begin{eqnarray}
\widehat{IC}_{\ell_1,\ell_2}^{(JWC)} = 
\sum_{j=1}^{\mathcal{J}^m+1} c_N 
\left(
\widehat{IC}_{\ell_1,\ell_2}^{(G,J)}(j) - 
\frac{\overline{n}_G}{n_S} \widehat{IC}_{\ell_1,\ell_2}^{(WRC,J)}(j)
\right) ,
\end{eqnarray}
where $c_n$ is a constant that can be tuned for small sample precision, $\overline{n}_G = (N - G + 1)/G$ and $n_S = (N - S + 1)/S$. Similar to \citet{BV_2017} we use $c_n = 1$ and $S = 1$, thus $\overline{n}_S = N$. $G$ parameter can be optimized based on number of intraday observations, more details can be found in \citet{zhang2011}.

Note the estimator consists of two parts. The first is the averaged version of the realized covariance (\autoref{eq_QV_RC}) for a specific wavelet scale \textit{j} on a grid size of $\overline{n} = N/G$: 
\begin{eqnarray}
\widehat{IC}_{\ell_1,\ell_2}^{(G,J)}(j) = \frac{1}{G} \sum_{g=1}^{G} \sum_{k=1}^N \mathcal{W}_{j,k}^{\ell_1} \mathcal{W}_{j,k}^{\ell_2},
\end{eqnarray}
where the wavelet coefficients are estimated on the jump-adjusted process $\Delta Y_{t,\ell}^{(J)} = (\Delta_1 Y_{t,\ell}^{(J)}, \ldots, \Delta_N Y_{t,\ell}^{(J)})$. 

The second part represents the part of the estimator from \autoref{eq_QV_RC} corresponding to a wavelet scale \textit{j}: 
\begin{eqnarray}
\widehat{IC}_{\ell_1,\ell_2}^{(WRC,J)}(j) = \sum_{k=1}^N \mathcal{W}_{j,k}^{\ell_1} \mathcal{W}_{j,k}^{\ell_2},
\end{eqnarray}

As discussed in \citet{BV_2017}, the estimator converges in probability to the integrated covariance $\widehat{IC}_{\ell_1,\ell_2}^{(JWC)} \overset{p}{\to} IC_{\ell_1,\ell_2}$. For the $\ell_1=\ell_2$, we obtain integrated variances, and hence the final estimate of the full integrated covariance matrix: 
\begin{eqnarray}
\label{eq_IC_JWC_matrix}
\widehat{\textbf{IC}}^{(JWC)} =
\begin{pmatrix}
\widehat{IC}_{\ell_1,\ell_1}^{(JWC)} & \widehat{IC}_{\ell_1,\ell_2}^{(JWC)} \\
\widehat{IC}_{\ell_2,\ell_1}^{(JWC)} & \widehat{IC}_{\ell_2,\ell_2}^{(JWC)}
\end{pmatrix}.
\end{eqnarray}

\section{The role of co-jumps in the U.S. and European interest rate futures markets}

The primary aim of this work is to shed light on the role of jumps and mainly co-jumps in the U.S. and European interest rate futures markets. In the subsequent sections, we will use the wavelet based techniques outlined above to identify co-jumps and discuss their role in the covariation of the yield curve first. Then, we will examine impact of news announcements on the co-jumps. Having high frequency data about both interest rate markets, we will also aim to document similarities and differences in the role of co-jumps for the two distinct markets.

\subsection{Exact co-jump detection}

A last step in detecting co-jumps is to consider bootstraping due to unknown distribution of the estimates. Bootstrapping of realized measures can significantly improve the finite sample properties of the jump \citep{dovonon2014bootstrapping} and co-jump tests. 

The main effort in the estimation is to recover co-jumps from the observed, noisy data. To do so, we can compare continuous covariation estimates with the estimate of whole quadratic covariation, and considering estimation error of both estimators, a standard Hausman-type test statistics can be constructed to test the null hypothesis of no jumps and co-jumps at a given day. In a univariate setting, \cite{barunikjwg} bootstrap this type of statistic for jump detection, \cite{BV_2017} further extend the strategy to co-jumps.\footnote{Technical details are given in \ref{sec:bootstrap}.}

Being able to identify days that have significant co-jump components, next step is to determine whether the reason why the null hypothesis of no co-jumps was rejected is because of the presence of co-jump(s) or, alternatively, because of the occurrence of large idiosyncratic (disjoint) jump(s). \cite{gnabo2014system} show that large idiosyncratic jumps may inflate the test statistic, and thus, co-jumps may be falsely detected. Therefore, there are basically two possible reasons why the null hypothesis was rejected:
\begin{enumerate}
\item Co-jumps: $t\in[0,T]$: $\Delta_i J_{t,\ell_1} \Delta_i J_{t,\ell_2} \neq 0$, i.e., the process is not exactly zero.
\item Disjoint jumps: $t\in[0,T]$: the processes $\Delta_i J_{t,\ell_1}$ and $\Delta_i J_{t,\ell_2}$ are not both zero (at least one of them), but $\Delta_i J_{t,\ell_1} \Delta_i J_{t,\ell_2} = 0$.
\end{enumerate}

An advantage of our approach is that the exact jump position is obtained by the wavelet analysis; hence, we can successfully eliminate the false co-jump situation caused by high idiosyncratic jump(s). Furthermore, because we know the directions of the jumps, we can distinguish between co-jumps that occur with jumps of the same or different direction on day $t$.

\subsection{Co-jump variation}

We begin the empirical analysis with estimation of covariance matrix for all 2 year - 5 year, 2 year - 10 year, 2 year - 30 year, 5 year - 10 year, 5 year - 30 year, and 10 year - 30 year pairs for both U.S. and European interest rate futures markets, and decompose the covariance into continuous part and co-jumps variation using the jump wavelet covariance estimator.   

\begin{table}[!h]
\footnotesize
\centering
\begin{tabular}{rrrrrrr}
\toprule
& \multicolumn{6}{c}{U.S.}\\
\cmidrule{2-7}{}
  & 2y - 5y & 2y - 10y & 2y - 30y & 5y - 10y & 5y - 30y & 10y - 30y \\
\cmidrule{2-7}
  Days with CJ$\ne0$ & 235 & 201 & 163 & 444 & 365 & 347 \\ 
  QV & 0.00379 & 0.00508 & 0.00689 & 0.01771 & 0.02579 & 0.04478 \\ 
  \% CJ/QV & 3.97 & 3.41 & 3.32 & 5.86 & 7.11 & 4.32 \\ 
  \cmidrule{2-7}
& \multicolumn{6}{c}{Europe}\\
\cmidrule{2-7}
   & 2y - 5y & 2y - 10y & 2y - 30y & 5y - 10y & 5y - 30y & 10y - 30y \\ 
\cmidrule{2-7}
    Days with CJ$\ne0$ & 144 & 102 & 60 & 217 & 145 & 240 \\ 
  QV & 0.00327 & 0.00430 & 0.00587 & 0.01527 & 0.02244 & 0.04807 \\ 
  \% CJ/QV & 1.84 & 1.40 & 0.87 & 2.31 & 1.58 & 2.02 \\
  \bottomrule
\end{tabular}
\caption{Co-jump statistics: Number of days with co-jumps, quadratic covariation (QV), and the average daily ratio of co-jump variation to quadratic covariation (\% CJ/QV).} 
\label{tab:cojumps}
\end{table}

Table \ref{tab:cojumps} documents behavior of co-jumps at both markets. We document substantially higher co-jumps activity in the U.S. market in comparison to European market, since number of days with significant co-jumps as well as ratio of co-jump variation to total variation showing role of the co-jumps in the covariance is more than doubled in most pairs.

The longer end of the yield curve with  5 year - 10 year and 5 year - 30 year pairs seem to be most influenced in both markets. In the U.S. market, significant co-jumps were estimated in 444 and 365 out of total 2705 days for the two pairs respectively. In the European market, we document highest jump activity in the 5 year - 10 year and 10 year - 30 year pairs, specifically, 217 and 240 out of total 2780 trading days. Looking at the shorter end of the curve, 2 year contract co-jumps mostly with the 5 year contact, and co-jumps activity slightly decreases with increasing maturity being lowest for the 2 year - 30 year pair logically.

Table \ref{tab:cojumps} further documents role of the co-jumps in the total covariation. While we document approximately twice larger role of co-jumps in the U.S. market in comparison to European market, co-jumps play also twice larger role between longer end of the curves in comparison to shorter end.

\subsection{Time dynamics of co-jump variation}

\begin{figure}[!h]
\centering
\includegraphics[width=\textwidth]{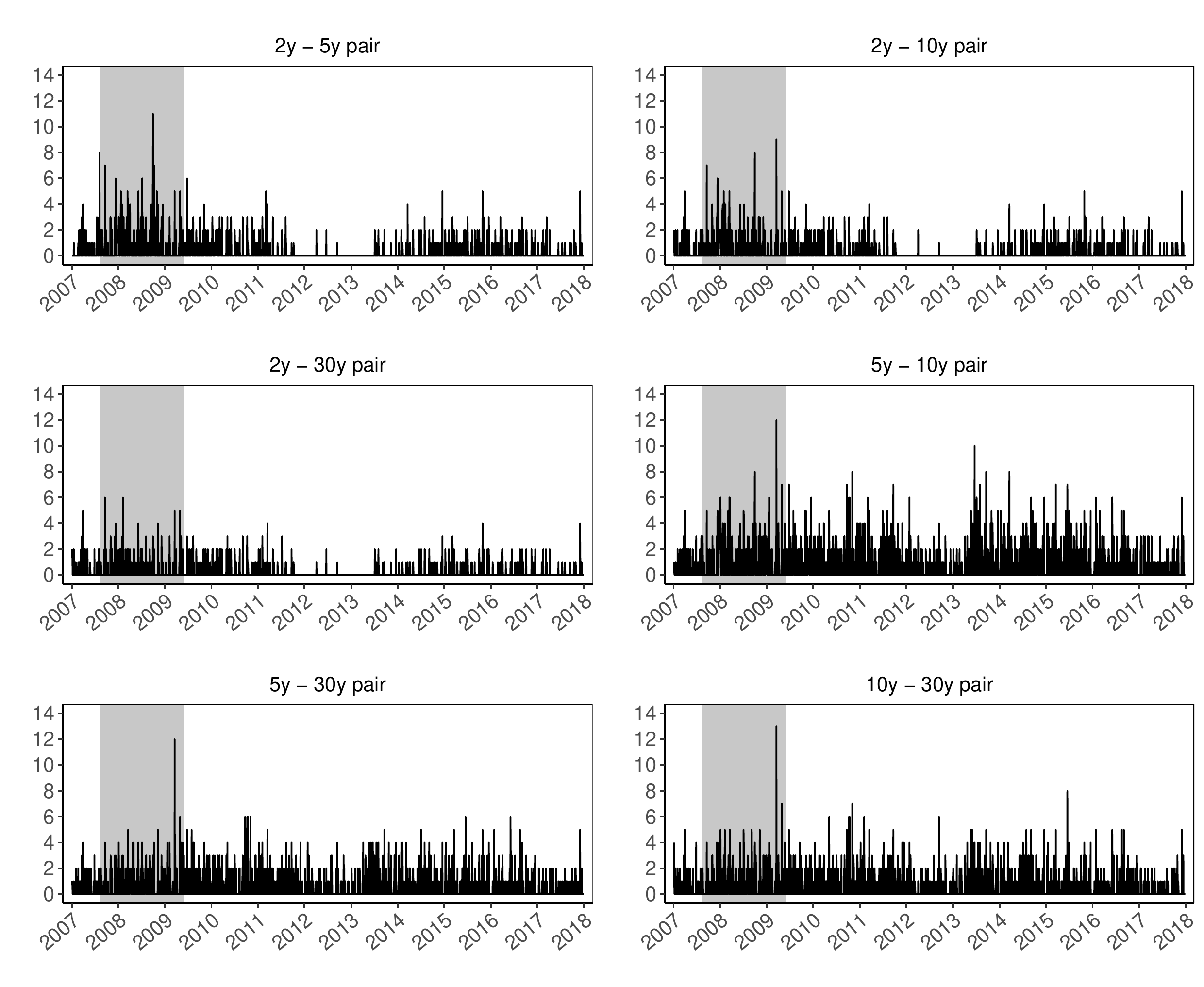}
\caption{Number of co-jumps for all U.S. pairs during the period under study.}
\label{fig:numberUS}
\end{figure}

\begin{figure}[!h]
\centering
\includegraphics[width=\textwidth]{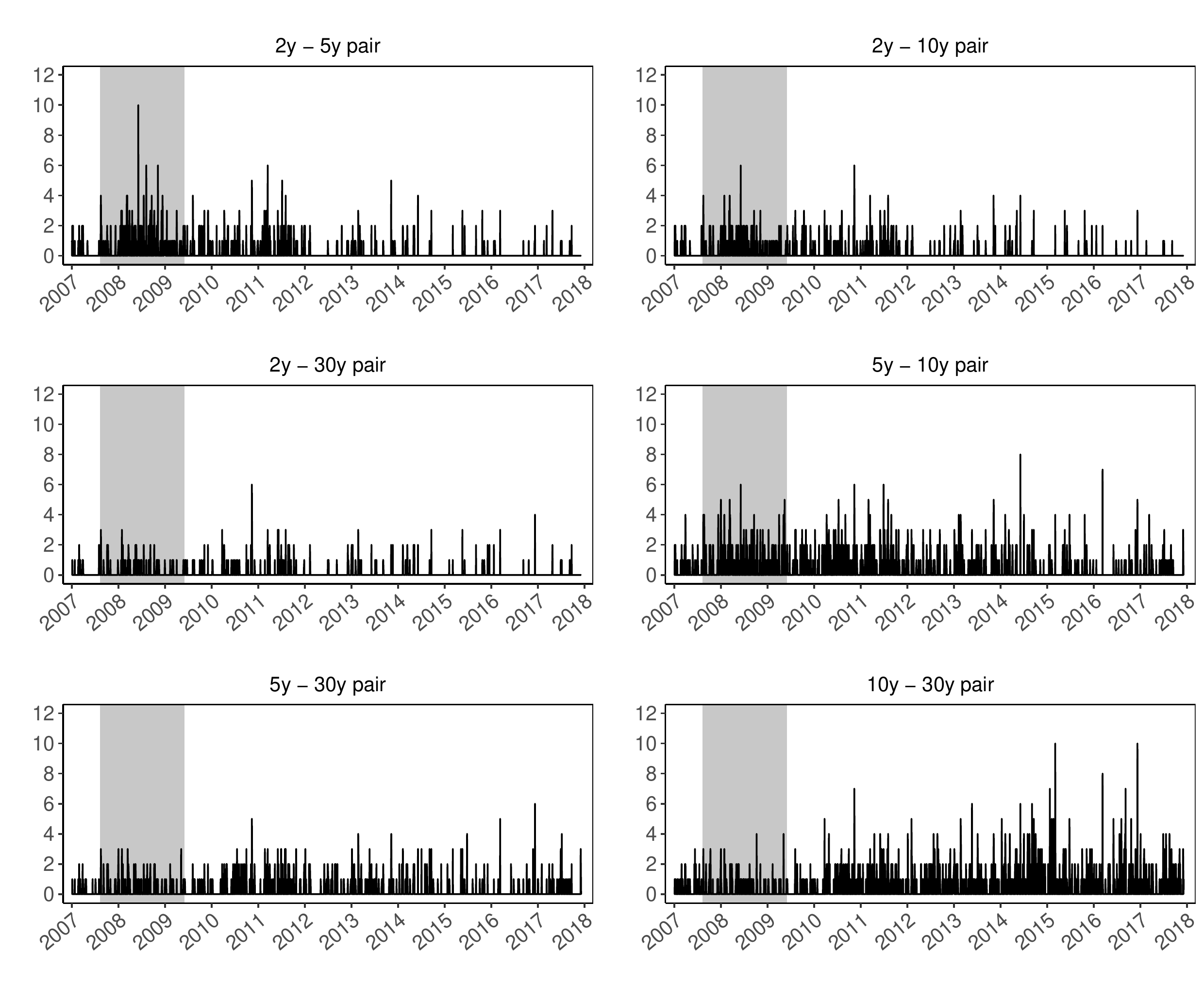}
\caption{Number of co-jumps for all European pairs during the period under study.}
\label{fig:numberEU}
\end{figure}

It is also interesting to look at the evolution of co-jumps in time. Figures \ref{fig:numberUS} and \ref{fig:numberEU} document number of co-jumps in the U.S. and European market during the estimation period respectively.

In addition to the previous results, Figure \ref{fig:numberUS} show that co-jumping activity of the U.S. 2 year contract with other contracts varies substantially in time. At the beginning of the sample activity was similar to the rest of pairs, while we document almost no co-jumps in the 2012 - 2013 period with slight rebound of the activity in the second half of the studied period. Possible explanation of this behavior is quantitative easing by FED. Activity of the rest of the pairs seems more stable over time. Situation with European contracts is similar, with lower activity for 2 year contract with other contracts in the second half of the studied period (Figure \ref{fig:numberEU}).

In addition, it is interesting to look at the evolution of the covariation and co-jumps. Figures \ref{fig:covUS1}, \ref{fig:covUS2}, \ref{fig:covEU1}, and \ref{fig:covEU2} in the \ref{app:figstabs} show the decomposition for the U.S. and European markets respectively. The evolution of the covariance over time reveals that all parts of the covariance matrix were exposed to increase during the financial crisis of 2007-2009 (highlighted in gray) for both markets. Further, European market was exposed to a large increase in correlations during 2011 in all pairs. This could be attributed to a Greek sovereign debt markets falling into severe stress before this period that caused to spread the crisis to other countries. Finally, the 10 year - 30 year pair documents rise during 2015 that could be attributed to a recovery from the crisis.

\subsection{Impact of co-jumps on correlation structure}

Having precise decomposition of  the quadratic variation, it is tempting to see how co-jumps impact the correlation structure of the yield curve. Total correlation defined as quadratic covariance normalized by volatilities of the two processes will be decreasing with increasing impact of idiosyncratic jumps, while increasing with the presence of co-jumps as discussed by \citep{BV_2017}. To study the influence of co-jumps on the correlation structure, we compare the correlations estimated using quadratic variation (containing all information) and correlation estimated using integrated covariance (excluding jumps and co-jumps). Estimates are available in the Table \ref{tab:corrs} in the \ref{app:figstabs}. If the two will be statistically indistinguishable, then co-jumps will not play significant role in the correlation structure. 

Hence, we aim to test the null hypothesis $H_0: \widehat{\text{corr}}_T^{QV} - \widehat{\text{corr}}_T^{IC} = 0$ by running a simple regression
\begin{equation}
\widehat{\text{corr}}_T^{QV} =\alpha + \beta \widehat{\text{corr}}_T^{IC} + \epsilon_T,
\end{equation}
with i.i.d. error with constant variance. In case $\alpha=0$ and $\beta=1$ jointly, we are not able to reject the equality of correlations with and without co-jumps. Furthermore, we pay special attention to constant coefficient $\alpha$ from the regression since a positive $\alpha$ would imply that the occurence of co-jumps plays an important role in total correlations. Conversly, negative $\alpha$ would imply that idiosyncratic jumps have a larger impact on total correlations than co-jumps.  

\begin{table}[!ht]
\footnotesize
\centering
\begin{tabular}{rrrrrrr}
  \toprule
  & \multicolumn{6}{c}{The U.S. market}\\
\cmidrule{1-7}
 & 2y - 5y & 2y - 10y & 2y - 30y & 5y - 10y & 5y - 30y & 10y - 30y \\
  \cmidrule{2-7}
$\alpha$ & 0.193 & 0.195 & 0.232 & 0.846 & 0.543 & 0.753 \\ 
  $\beta$ & 0.786 & 0.758 & 0.632 & 0.115 & 0.369 & 0.186 \\ 
   $R^2$ & 0.629 & 0.614 & 0.494 & 0.100 & 0.262 & 0.153 \\ 	
\cmidrule{1-7}
   & \multicolumn{6}{c}{European market}\\
\cmidrule{1-7}
& 2y - 5y & 2y - 10y & 2y - 30y & 5y - 10y & 5y - 30y & 10y - 30y \\
 \cmidrule{2-7}
 $\alpha$ & 0.242 & 0.184 & 0.189 & 0.734 & 0.275 & 0.319 \\ 
  $\beta$ & 0.721 & 0.758 & 0.683 & 0.204 & 0.683 & 0.666 \\ 
   $R^2$ & 0.575 & 0.589 & 0.518 & 0.140 & 0.589 & 0.662 \\ 
\bottomrule
\end{tabular}
\caption{The table shows the estimated coefficients from the regression $\widehat{\text{corr}}_T^{QV} =\alpha + \beta \widehat{\text{corr}}_T^{IC} + \epsilon_T$. The null hypothesis of joint insignificant difference of $\alpha$ from zero and $\beta$ from one is rejected in all cases using Wald test with heteroscedasticity consistent White's covariance estimator.}
\label{tab:impact}
\end{table}

Table \ref{tab:impact} shows the estimation results for both markets and reveals that co-jumps seem to be a significant part of total correlations. Whereas both U.S. and European markets show  similar results except the 5 year - 10 year and 10 year - 30 year futures, co-jumps impact correlation structure much heavily in the long end of the yield curves, although short end of the curve also shows significant increase of correlations due to co-jumps.

\subsection{Effect of news announcements on co-jumps}

In the previous section, we have documented significant role of co-jumps in the U.S. and European interest rates futures market, with long end of the curves having more co-jumps activity. Next, we would like to connect this activity to the news announcements.

Scheduled macroeconomic news announcement causing jumps and co-jumps of the yield curve imply significant unexpected changes in the economic behavior or monetary policy. \citet{Balduzzi_1997}, \citet{Balduzzi_2001}, \citet{piazzesi_2005}, \citet{J_Dungey_2009}, \citet{lahaye_2011} and \citet{CJ_Dungey_2012} document a negative relationship between news surprises and bond returns. The strongest effect is found when news about non-farm payrolls are released followed by output, and inflation announcements. In addition, the size of the effect differs across maturities and the effect is increasing with the bond's maturity. This is consistent with the higher volatility of bonds at the long end of the yield curve.

\citet{J_Dungey_2009} further document that two thirds of the co-jumps in the yield curve structure coincide with the scheduled US news release. Moreover, according to \citet{CJ_Dungey_2012} the probability of co-jump occurrence increases with the scheduled macroeconomic news announcement. However, the research is still limited and search for the sources of the co-jumps continues since not all news surprises generate co-jump and conversely not all co-jumps in the term structure are connected with the news surprise. 

\begin{figure}[!h]
\centering
\includegraphics[width=\textwidth]{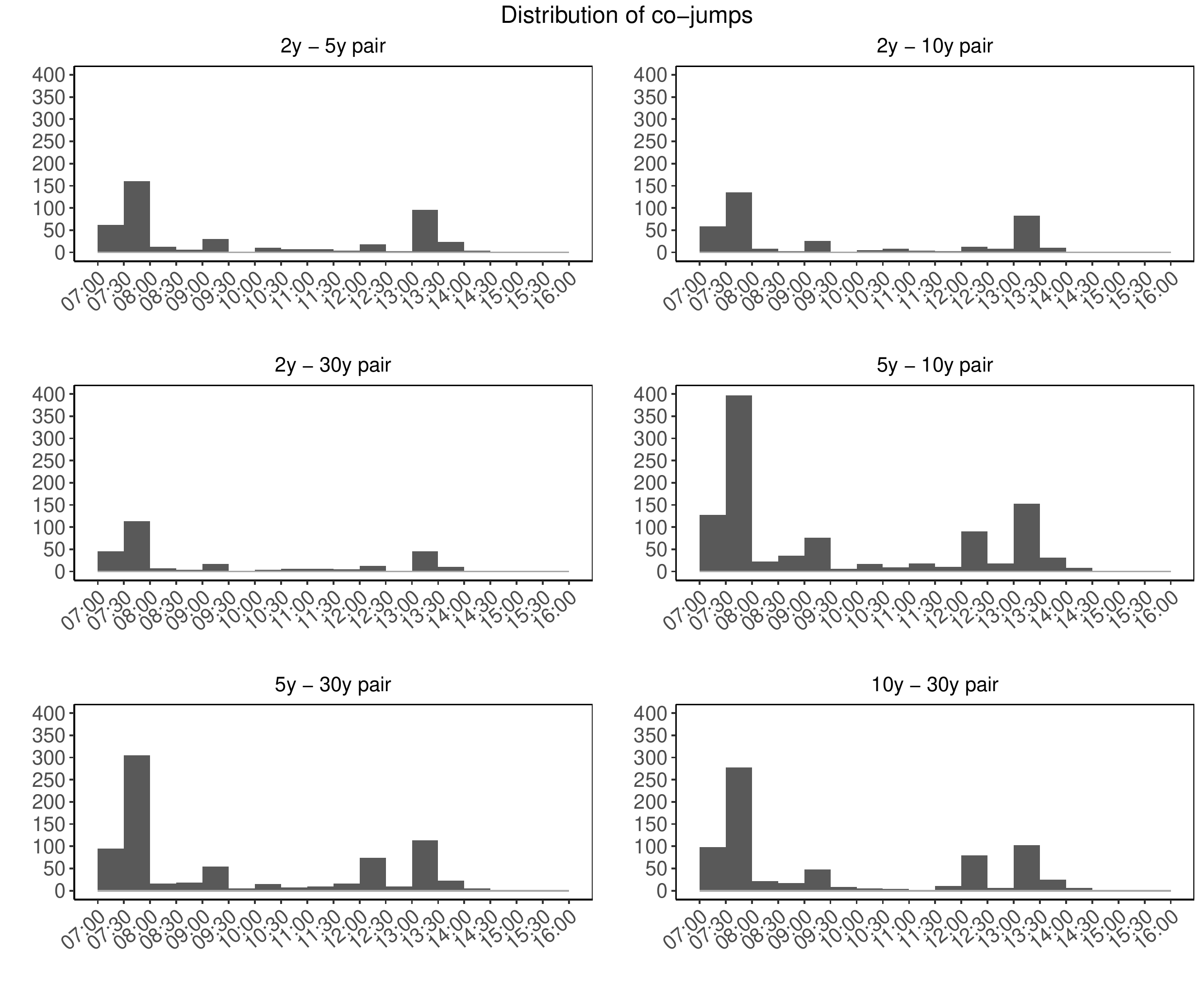}
\caption{Distribution of the U.S. interest rate futures co-jumps during the trading session.}
\label{fig:histUS}
\end{figure}

\begin{figure}[!h]
\centering
\includegraphics[width=\textwidth]{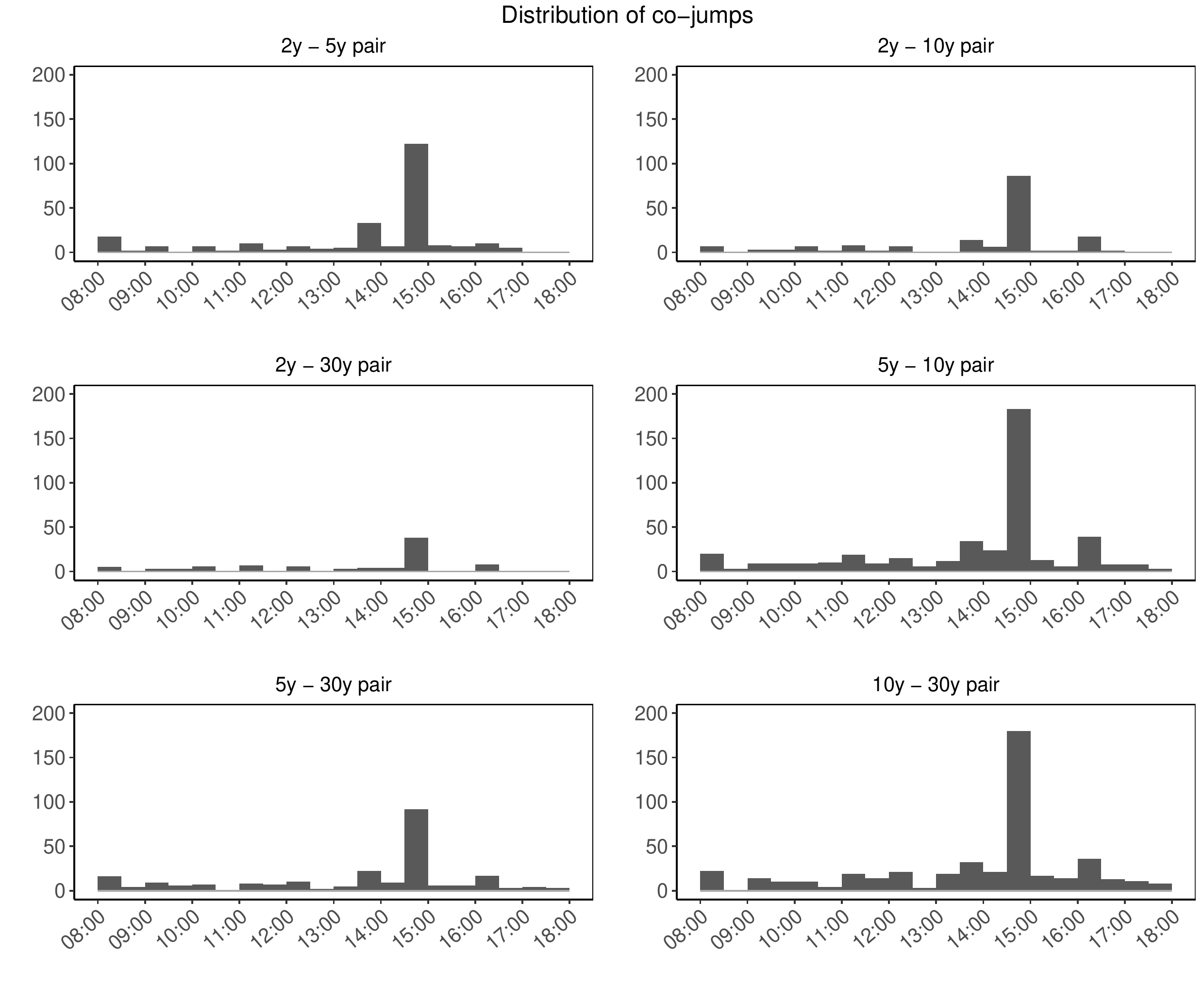}
\caption{Distribution of the European interest rate futures co-jumps during the trading session.}
\label{fig:histEU}
\end{figure}

Figure \ref{fig:histUS} show distribution of co-jumps during the trading session. Note the largest amount of co-jumps between U.S. interest rate futures happens during 7:30-8:00 CST linked to higher trading volume. Second peak is between 13:00 and 13:30 CST, around the FOMC press releases scheduled at 13:00 CST. Similarly in the European market, as shown by Figure \ref{fig:histEU}, most of the co-jump activity is at 14:30 - 15:00 CEST, around the ECB announcements. The decision on the key rate is published via press release at 13:45 CET and is further  explained at a press conference starting at 14:30 CET, see \ref{sec:eudata}.

To test the hypothesis that co-jumps are connected with news announcements, we estimate a simple regression in which we explain probability of co-jump occurrence by a scheduled press release. In case of the U.S. market, we use FOMC press releases, and take 30 minute time window following the announcement, with 103 FOMC announcement days in the data. In the European market, we use ECB press releases, and take 15 minute window after the press release, and 30 minute window after the conference as suggested by \citet{CJ_ECB_2014}, with 113 ECB announcement days. We run a regression
\begin{equation}
\Pr\{\widehat{CJ}_T\ne0|\text{news}_T\} = \Lambda(\theta),
\end{equation}
where $\Lambda(\theta)$ is a logistic function with $\theta=\beta_0+\beta_1 \text{news}_T$, modeling probability of co-jump conditional on the news. $H_0: \beta_1=0$ tests the hypothesis that news do not influence co-jumps.

Table \ref{tab:logit} summarizes the results from estimation. Since both coefficients are significantly different from zero, probability of jump occurrence is different from 0.5. Negative $\beta_0$ points to probability of co-jump occuring in days without scheduled announcement is below 0.5. The more negative value of the $\beta_0$, the smaller the probability is. Positive significant $\beta_1$ identifies increase in probability of co-jump in case of the news. 

\begin{table}[!h]
\footnotesize
\centering
\begin{tabular}{rrrrrrr}
  \toprule
  & \multicolumn{6}{c}{FOMC and the U.S. market}\\
\cmidrule{1-7}
 & 2y - 5y & 2y - 10y & 2y - 30y & 5y - 10y & 5y - 30y & 10y - 30y \\
  \cmidrule{2-7}
$\beta_0$ & -5.782 & -5.782 & -6.253 & -5.024 & -5.558 & -5.462  \\ 
  $\beta_1$ & 5.203 & 4.985 & 4.700 & 4.570 & 4.893 & 4.665  \\ 
  $R^2$ & 0.469 & 0.433 & 0.366 & 0.380 & 0.425 & 0.389  \\ 
\cmidrule{2-7}
	& \multicolumn{2}{c}{2y - 5y - 10y} & \multicolumn{2}{c}{5y - 10y - 30y} & \multicolumn{2}{c}{2y - 5y - 10y - 30y} \\
	\cmidrule{2-7}
$\beta_0$ &  \multicolumn{2}{c}{-6.764} & \multicolumn{2}{c}{-6.070} & \multicolumn{2}{c}{-5.782} \\ 
  $\beta_1$ & \multicolumn{2}{c}{4.829} & \multicolumn{2}{c}{4.932} & \multicolumn{2}{c}{4.644} \\ 
  $R^2$ &  \multicolumn{2}{c}{0.364} & \multicolumn{2}{c}{0.412} & \multicolumn{2}{c}{0.375} \\ 
\cmidrule{1-7}
   & \multicolumn{6}{c}{ECB and European market}\\
\cmidrule{1-7}
& 2y - 5y & 2y - 10y & 2y - 30y & 5y - 10y & 5y - 30y & 10y - 30y \\
 \cmidrule{2-7}
$\beta_0$ & -3.997 & -4.208 & -4.787 & -3.424 & -3.822 & -3.318  \\ 
  $\beta_1$ & 3.081 & 2.609 & 2.015 & 2.336 & 1.723 & 1.658  \\ 
  $R^2$ & 0.157 & 0.099 & 0.044 & 0.075 & 0.032 & 0.030  \\ 
\cmidrule{2-7}
		& \multicolumn{2}{c}{2y - 5y - 10y} & \multicolumn{2}{c}{5y - 10y - 30y} & \multicolumn{2}{c}{2y - 5y - 10y - 30y} \\
		\cmidrule{2-7} 
$\beta_0$ &  \multicolumn{2}{c}{-4.234} & \multicolumn{2}{c}{-4.018} & \multicolumn{2}{c}{-4.989} \\ 
  $\beta_1$ & \multicolumn{2}{c}{2.510} & \multicolumn{2}{c}{1.831} & \multicolumn{2}{c}{1.863} \\ 
  $R^2$ &  \multicolumn{2}{c}{0.089} & \multicolumn{2}{c}{0.037} & \multicolumn{2}{c}{ 0.034} \\ 
\bottomrule
\end{tabular}
\caption{Co-jumps and news announcement. All reported coefficients are statistically different from zero at 1\% level of significance.}
\label{tab:logit}
\end{table}

Looking at the results from the U.S. market, we can see significant impact of the scheduled FOMC press releases on co-jump occurrence in the 30 minute window after the news release with highest impact on 2 year - 5 year pair, and slightly declining in maturities.

As for the European market, ECB news releases seem to have much weaker influence on the co-jumps influence, with highest influence at 2 year - 5 year pair again, but dramatically lower explanatory power in comparison to the U.S. market.

Importantly, we also run the models explaining probability of the co-jumps in three contracts, and finally in the whole curve. Table \ref{tab:logit} documents significant impact of the FOMC announcements on the U.S. interest rate futures even in the case of multi-jumps. Similar, but weaker efect is also documented in the European market. This means that news announcements in both markets have significant impact on the whole yield curve.

\subsection{Level shift vs. rotation of the yield curve}
\label{theory_rot_shift}

Connecting the news with co-jumps, it is interesting to deeper look at the reaction of the yield curve on the news. Generally, there are two possible reactions of the yield curve to market changes - level shift and rotation of the yield curve \citep{CJ_ECB_2014,J_Dungey_2009,ellingsen_2001}. With this respect, literature has studied a monetary policy rule proposed by \citet{Taylor_1993}
\begin{eqnarray}
\label{eq_taylor}
\Delta r_t &=& \lambda \Delta \pi_t + (1-\lambda)\Delta y_t, \hspace{0.5cm} \lambda \in [0,1],
\end{eqnarray}
which represents reaction function of a central bank. \autoref{eq_taylor} explains the relationship between the change of interest rate $\Delta r_t$ as a function of change of inflation expectations $\Delta \pi_t$ and change in output $\Delta y_t$, where $\lambda$ is the preference parameter of central bank assigning the relative importance to each of the macroeconomic variables. The interpretation of the Taylor rule is straightforward - with increasing inflation or increasing output, central banks proportionally to $\lambda$, $1-\lambda$ respectively, increase the interest rate.

Two possible sources of policy surprise can be identified from equation \autoref{eq_taylor} \citep{CJ_ECB_2014,ellingsen_2001}. First, any macroeconomic news captured by $\Delta \pi_t$ and $\Delta y_t$ that impacts the bond prices should affect all maturities and co-jump across the whole structure should be observed - level shift of the yield curve (endogenous monetary policy). Second, according to liquidity preference theory of the long term structure the long term bond returns are higher as they reflect higher liquidity risk premium given their longer maturity. Thus, any change in the preference parameter $\lambda$ reflecting the changes of the central bank monetary policy preferences causes short and long end of the yield curve to move in opposite direction - rotation of yield curve (exogenous monetary policy). 

With information asymmetry about the shocks to the economy the market interest rate will be positively correlated with the interest rate set by a central bank. Therefore, positive inflationary shock, representing the increase in current inflation and output (monetary tightening), increases inflation expectations across all maturities. Thus, the entire yield curve shift upwards. For a negative inflationary shock, representing the decrease in current inflation and output (monetary expansion), the yield curve shifts downwards. 

The response of the interest rate to unexpected change in the central bank preferences is captured by the parameter $\lambda$. After the shock hits the economy, market participants expect central bank to behave according to the \autoref{eq_taylor}. When central bank responds unexpectedly it leads market participants to permanently change their expectations about the preference parameter $\lambda$. Thus, if central bank sets higher interest rate than expected, the market will adjust their expectations thinking central bank has become more inflation averse - decrease $\lambda$. This would lead to rising short-term interest rates but falling long-term rates. 

Increasing $\lambda$ implies increasing preference of stable output over inflation close to target. This results in higher shock persistence of the long term interest rate as the central bank rate is expected to deviate from the initial level for a longer period. Therefore, central banks with higher preference parameter will experience larger effect on long-term end of the yield curve.

Moreover, \citet{J_Dungey_2009} and \citet{Balduzzi_1997} remark that the trading activity on the short end of the yield curve might be higher, as there is higher liquidity on the market which attracts more speculators in comparison with the long term maturity markets. As a consequence the changes in monetary policy mainly influence the short end of the yield curve. Thus, the jumps might occur more often in the short end of the yield curve. Iinterestingly, this is not confirmed in our data as discussed earlier.

\subsection{The level shifts and rotations of the yield curve}
\label{test_rot_shift}

The ability to identify days where the null hypothesis is rejected because of the presence of idiosyncratic jumps and significant co-jumps helps us in the analysis of level shifts and rotations of the term structure. Since we are able to identify the days where only co-jumps play significant role, we are able to build on \citet{CJ_ECB_2014}'s level shifts and rotation hypothesis and study the direction of the co-jump within the intraday directly.

\begin{table}[!ht]
\scriptsize
\centering
\begin{tabular}{lrrrrrrrrrrr}
  \toprule
  &\multicolumn{2}{c}{R}& & \multicolumn{2}{c}{$\uparrow$ LS}& & \multicolumn{2}{c}{$\downarrow$ LS}& & \multicolumn{2}{c}{Days CJ$\ne$0} \\
 \cmidrule{2-3} \cmidrule{5-6} \cmidrule{8-9} \cmidrule{11-12}
 & \# & \% & & \# & \% & & \# & \% & & \# & \%  \\ 
\midrule
\textbf{FOCM}&\multicolumn{11}{c}{announcement days}\\
\midrule
  2y - 5y 				& 1 & 0.94 & & 19 & 17.92 & & 16 & 15.09 & & 37 & 34.91 \\ 
  2y - 10y 				& 1 & 0.94 & & 17 & 16.04 & & 14 & 13.21 & & 32 & 30.19 \\ 
  2y - 30y 				& 1 & 0.94 & & 9 & 8.49 & & 7 & 6.60 & & 18 & 16.98 \\ 
  5y - 10y 				& 0 & 0    & & 20 & 18.87 & & 20 & 18.87 & & 40 & 37.74 \\ 
  5y - 30y 				& 1 & 0.94 & & 15 & 14.15 & & 19 & 17.92 & & 35 & 33.02 \\ 
  10y - 30y 			& 1 & 0.94 & & 15 & 14.15 & & 16 & 15.09 & & 32 & 30.19 \\ 
  2y - 5y - 10y			& 2 & 1.89 & & 12 & 11.32 & & 11 & 10.38 & & 25 & 23.58 \\ 
  5y - 10y - 30y 		& 3 & 2.83 & & 12 & 11.32 & & 10 & 9.43 & & 25 & 23.58 \\ 
  2y - 5y - 10y - 30y 	& 2 & 1.88 & & 6 & 5.66 & & 5 & 4.72 & & 13 & 12.26 \\ 
  \midrule 
  &\multicolumn{11}{c}{non - announcement days} \\
  \midrule 
  2y - 5y 				& 0 & 0 & & 5 & 0.19 & & 3 & 0.12 & & 8 & 0.31 \\ 
  2y - 10y 				& 0 & 0 & & 4 & 0.15 & & 4 & 0.15 & & 8 & 0.31 \\ 
  2y - 30y 				& 0 & 0 & & 3 & 0.12 & & 2 & 0.08& & 5 & 0.19 \\ 
  5y - 10y 				& 0 & 0 & & 6 & 0.23 & & 11 & 0.42 & & 17 & 0.65 \\ 
  5y - 30y 				& 0 & 0 & & 4 & 0.15 & & 6 & 0.23 & & 10 & 0.38 \\ 
  10y - 30y 			& 0 & 0 & & 6 & 0.23 & & 5 & 0.19 & & 11 & 0.42 \\ 
  2y - 5y - 10y 		& 0 & 0 & & 3 & 0.12 & & 3 & 0.12 & & 6 & 0.23 \\ 
  5y - 10y - 30y 		& 0 & 0 & & 4 & 0.15 & & 4 & 0.15 & & 8 & 0.31 \\ 
  2y - 5y - 10y - 30y 	& 0 & 0 & & 2 & 0.08 & & 1 & 0.04 & & 3 & 0.12 \\ 
  \midrule 
  \textbf{ECB} &\multicolumn{11}{c}{announcements days} \\
  \midrule 
  2y - 5y 				& 0 & 0 & & 16 & 13.45 & & 18 & 15.13&  & 34 & 28.57 \\ 
  2y - 10y 				& 0 & 0 & & 9 & 7.56 & & 11 & 9.24 & & 20 & 16.81 \\ 
  2y - 30y 				& 0 & 0 & & 1 & 0.84 & & 6 & 5.04 & & 7 & 5.88 \\ 
  5y - 10y 				& 0 & 0 & & 12 & 10.08 & & 18 & 15.13 & & 30 & 25.21 \\ 
  5y - 30y 				& 0 & 0 & & 5 & 4.20 & & 8 & 6.72 & & 13 & 10.92 \\ 
  10y - 30y 			& 0 & 0 & & 8 & 6.72 & & 11 & 9.24 & & 19 & 15.97 \\ 
  2y - 5y - 10y 		& 0 & 0 & & 7 & 5.88 & & 11 & 9.24 & & 18 & 15.13 \\ 
  5y - 10y - 30y 		& 1 & 0.84 & & 5 & 4.20 & & 6 & 5.04 & & 12 & 10.08 \\ 
  2y - 5y - 10y - 30y 	& 0 & 0 & & 1 & 0.84 & & 4 & 3.36 & & 5 & 4.20 \\ 
  \midrule 
  &\multicolumn{11}{c}{non - announcements days} \\
  \midrule 
  2y - 5y 				& 0 & 0 & & 24 & 0.90 & & 24 & 0.90 & & 48 & 1.80 \\ 
  2y - 10y 				& 0 & 0 & & 20 & 0.75 & & 19 & 0.71 & & 39 & 1.47 \\ 
  2y - 30y 				& 0 & 0 & & 10 & 0.38 & & 12 & 0.45 & & 22 & 0.83 \\ 
  5y - 10y 				& 0 & 0 & & 41 & 1.54 & & 43 &   1.62 & & 84 & 3.16 \\ 
  5y - 30y 				& 0 & 0 & & 30 & 1.13 & & 27 &   1.01 & & 57 & 2.14 \\ 
  10y - 30y 			& 4 & 0.15 & & 50 & 1.88 & & 35 & 1.32 & & 93 & 3.49 \\ 
  2y - 5y - 10y 		& 3 & 0.11 & & 16 & 0.60 & & 19 & 0.71 & & 38 & 1.43 \\ 
  5y - 10y - 30y 		& 2 & 0.08 & & 25 & 0.94 & & 20 & 0.75 & & 47 & 1.77 \\ 
  2y - 5y - 10y - 30y 	& 0 & 0 & & 7 & 0.26 & & 11 &  0.41 & & 18 & 0.68 \\
   \bottomrule
\end{tabular}
\caption{Level shifts and rotations. R - rotation, LS - level shift}
\label{tab:rotation}
\end{table}

The level shift hypothesis can be formalized as a situation in which jumps of the same direction are identified simultaneously on the short end and long end of the yield curve. This co-jump estimated over [0 $\leq$ t $\leq$ T] from \autoref{eq_CJ_definition} leads to a parallel shift of the yield curve:
\begin{eqnarray}
\label{eq_hyp_shift}
\widehat{CJ}_{\ell_1,\ell_2}^{(i,t)} = \Delta_i J_{t,\ell_1} \Delta_i J_{t,\ell_2} > 0. 
\end{eqnarray}
Two types of parallel shifts can be present. Upward shift occurs if both jumps are positive $\Delta_i J_{t,\ell_1} > 0 $ $\Delta_i J_{t,\ell_2} > 0$ and vice versa if both jumps are negative $\Delta_i J_{t,\ell_1} < 0 $ $\Delta_i J_{t,\ell_2} < 0$ yield curve shifts downwards.

To the opposite, rotation of the yield curve occurs when both, short and long, ends of the yield curve simultaneously jump in the opposite direction, i.e. if the signs of jumps are of the opposite sign: 
\begin{eqnarray}
\label{eq_hyp_rot}
\widehat{CJ}_{\ell_1,\ell_2}^{(i,t)} = \Delta_i J_{t,\ell_1} \Delta_i J_{t,\ell_2} < 0. 
\end{eqnarray}

Table \ref{tab:rotation} show results for the level shifts and rotations identified from data using bootstrap. We focus on rotation and level shifts during the announcement days at both markets, but we also collect information about movement of the curve in a non-announcement days. 

Looking at the representative 2 year - 10 year pair, we identify 32 days with significant co-jumps in the sample of 103 FOMC press release days. Only one occasion happened to be a negative co-jump implying the market interpreted the announcement as a change in FED policy preferences. This co-jump is associated with March 18, 2008 FED decision to cut the federal fund rates by 0.75 percentage points to further support the growth during the crisis. The rest of the identified co-jumps was positive indicating 31 days at which the yield significantly shifted its level. These days are viewed by market as a policy surprise informing about the current and future state of the US economy. We document 17 positive policy surprise days resulting in upward shift, and 14 negative inflationary shocks resulting in downward shift

To emphasize the importance of co-jumps, we also collect frequency of rotations and level shifts on non-announcement days, that is the rest of the sample. To ensure consistency of the results, the co-jump activity is examined during the sample period as in the announcement days. Looking at the 2 year - 10 year pair, we only document only 8 days out of the 2602 non-announcement days.

Looking at the rest of the maturities at the U.S. market, we can see a consistent pattern of the FOMC news impacting all pairs similarly. The most interesting is the result for all four contracts. We document 13 days in which the whole curve moves during the announcement. The curve rotates in two occasions, shifts upwards in six occasions, and shifts downwards in the five days. The rest of the non-announcement days sample shows only 3 out of 2602 days with significant shift of the curve not associated with news. 

Turning our attention to the European market, we can see 
the results are similar during the announcement days, with less movement at far end of the curve. The whole curve moves only 5 times during the announcement days. In contrast to the U.S. market, we document much higher number of non-announcement days suggesting that movement of the curve is also associated with other events, although the number of days is still very small, since the sample contains 2661 non-announcement days. The results are also consistent with \citet{CJ_ECB_2014} who report the similar results for the effect of ECB policy announcements on 2 year - 10 year German bond futures.

\section{Conclusion}

In this paper, we study the role of co-jumps in the U.S. and European interest rate futures markets using precise wavelet based localization techniques.

We document the importance of co-jumps in both markets, its time dynamics, and impact on the covariance structure of the interest rate market. We also document high probability of co-jump occurrence during the announcement days using the data about announcement days. Further, we show that approximately 30\% of the news are accompanied by level shift of the whole yield curve in both markets. On the contrary, only 1\% of announcements are interpreted by the market as an adjustment in preferences. 

For these reasons, quantitative models should take the role of jumps and co-jumps into consideration when evaluating the trade opportunities. Similarly, policy making entities should account for these discrepancies in yield curve as well as it should be in their best interest to set the appropriate monetary policy to ensure their targets are met. Last but not least, academic researches ought to take a keen interest in the subject as understanding the effect of monetary policy on real economy was always one of the topic attracting attention. 

\bibliography{biblio}
\bibliographystyle{chicago}

\clearpage
\newpage

\appendix
\section{Supplementary Tables and Figures}
\label{app:figstabs}

\begin{table}[!h]
\footnotesize
\centering
\begin{tabular}{rrrrrrr}
\toprule
& \multicolumn{6}{c}{U.S.}\\
\cmidrule{2-7}{}
  & 2y - 5y & 2y - 10y & 2y - 30y & 5y - 10y & 5y - 30y & 10y - 30y \\
\cmidrule{2-7}
Total correlation & 0.813 & 0.752 & 0.637 & 0.959 & 0.850 & 0.921 \\ 
  Continuous correlation & 0.731 & 0.682 & 0.590 & 0.894 & 0.783 & 0.853 \\ 
  Correlation difference & 0.054 & 0.048 & 0.036 & 0.059 & 0.059 & 0.065 \\ 
  \cmidrule{2-7}
& \multicolumn{6}{c}{Europe}\\
\cmidrule{2-7}
   & 2y - 5y & 2y - 10y & 2y - 30y & 5y - 10y & 5y - 30y & 10y - 30y \\ 
\cmidrule{2-7}
  Total correlation & 0.836 & 0.714 & 0.553 & 0.923 & 0.764 & 0.858 \\ 
  Continuous correlation & 0.782 & 0.670 & 0.517 & 0.859 & 0.677 & 0.771 \\ 
  Correlation difference & 0.039 & 0.035 & 0.036 & 0.056 & 0.079 & 0.068 \\ 
  \bottomrule
\end{tabular}
\caption{Median of the total correlation, continuous correlation, and their differences for both markets.} 
\label{tab:corrs}
\end{table}

\begin{figure}[!h]
\centering
\includegraphics[width=\textwidth]{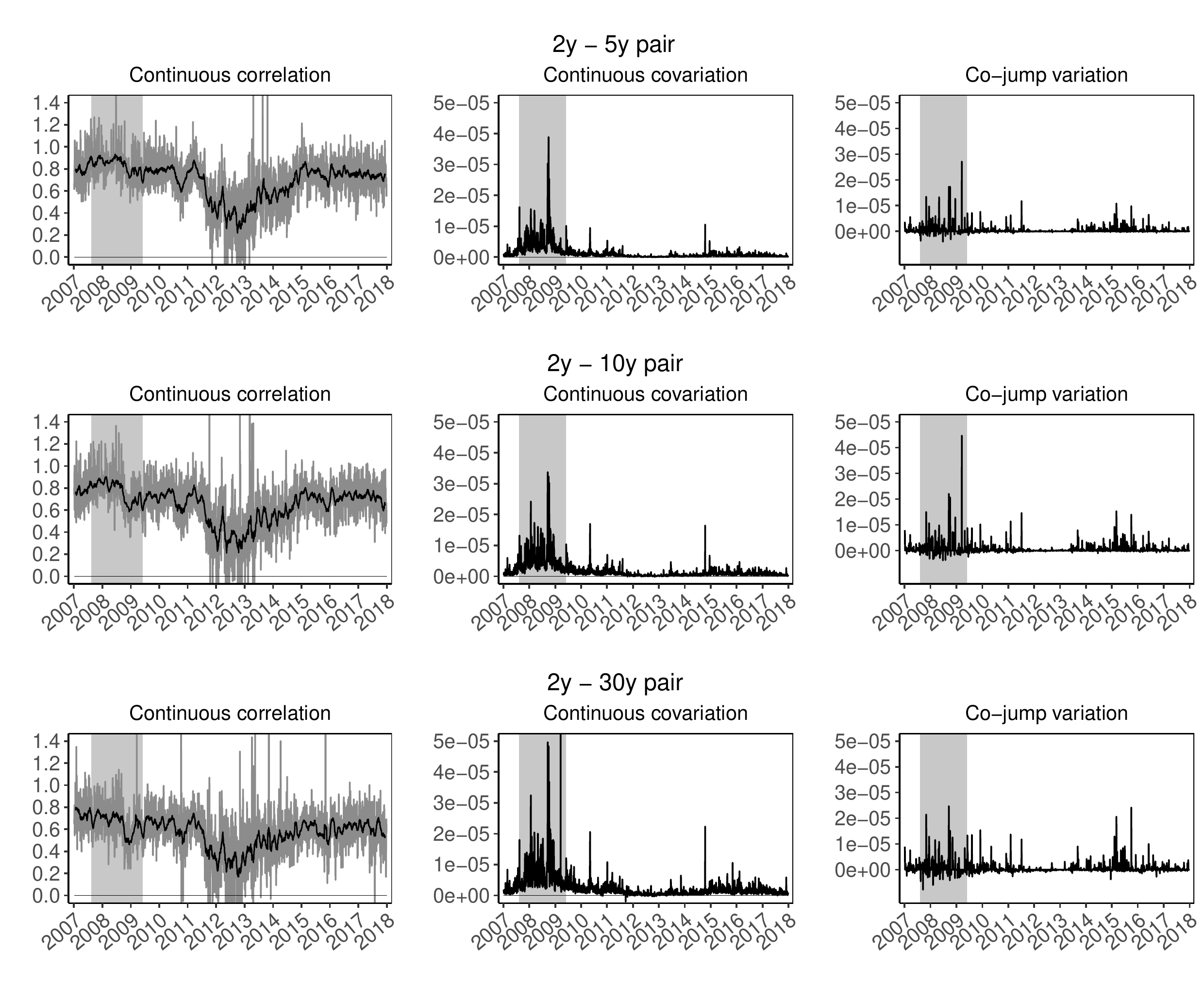}
\caption{U.S. interest rates market: Decomposition of the covariation for the 2 year - 5 year, 2 year - 10 year, and 2 year - 30 year pairs. Continuous correlation in gray with a 21-day moving average in black (left column), integrated covariance (middle column), and co-jumps (right column) estimated by jump wavelet covariance estimator. The 2007 -- 2008 crisis period is shaded.}
\label{fig:covUS1}
\end{figure}

\begin{figure}[!h]
\centering
\includegraphics[width=\textwidth]{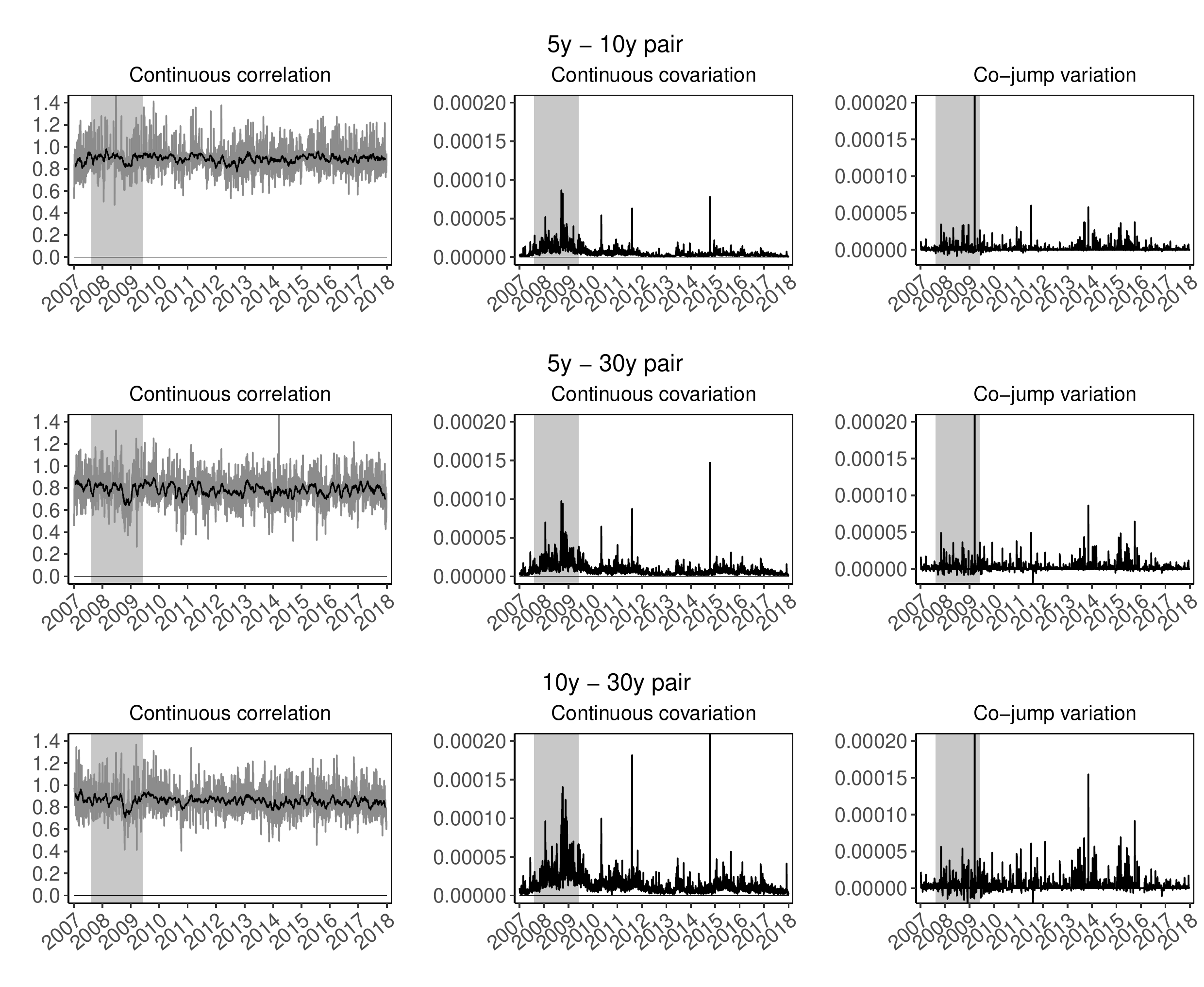}
\caption{U.S. interest rates market: Decomposition of the covariation for the 5 year - 10 year, 5 year - 30 year, and 10 year - 30 year pairs. Continuous correlation in gray with a 21-day moving average in black (left column), integrated covariance (middle column), and co-jumps (right column) estimated by jump wavelet covariance estimator. The 2007 -- 2008 crisis period is shaded.}
\label{fig:covUS2}
\end{figure}

\begin{figure}[!h]
\centering
\includegraphics[width=\textwidth]{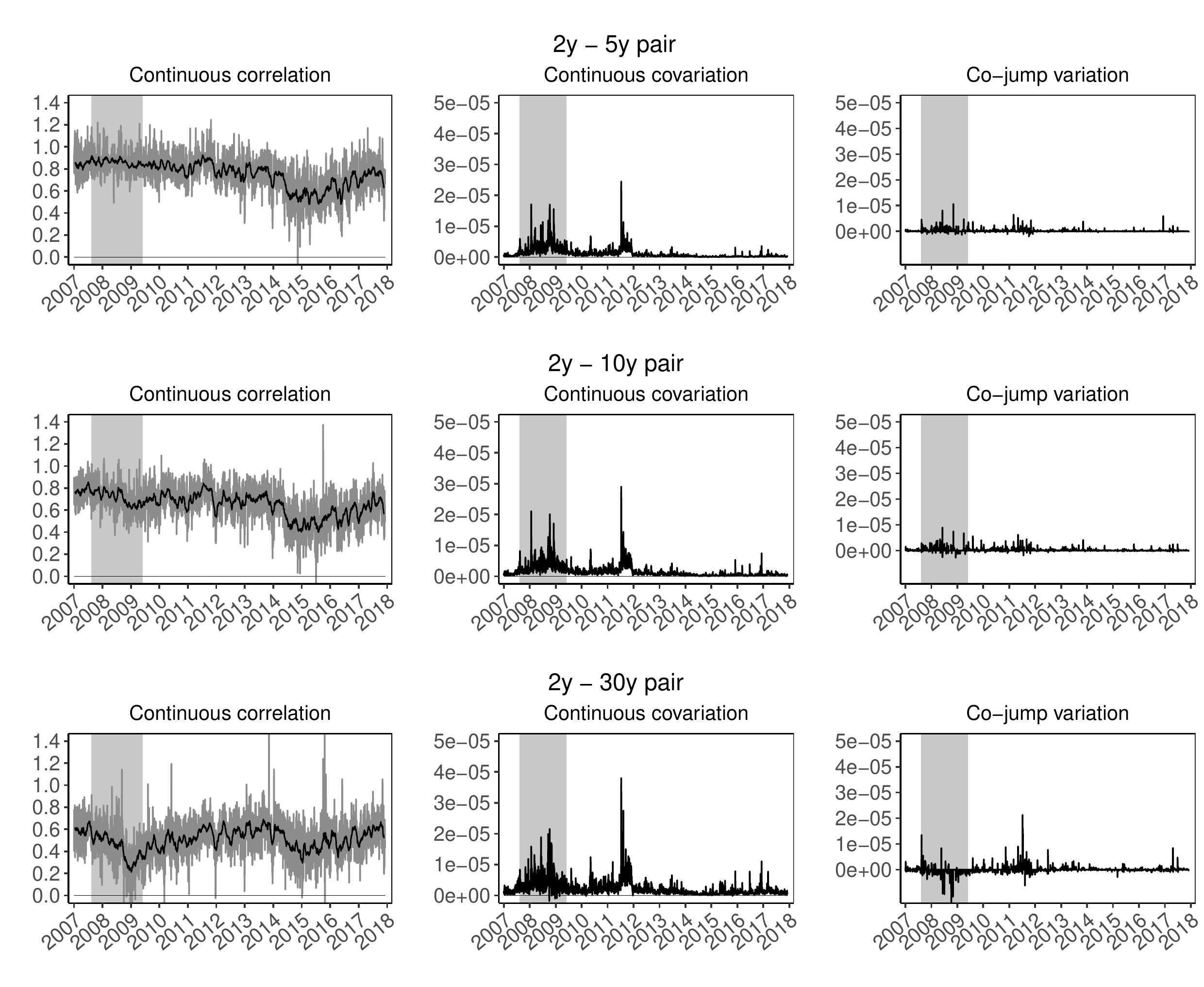}
\caption{European interest rates market: Decomposition of the covariation for the 2 year - 5 year, 2 year - 10 year, and 2 year - 30 year year pairs. Continuous correlation in gray with a 21-day moving average in black (left column), integrated covariance (middle column), and co-jumps (right column) estimated by jump wavelet covariance estimator. The 2007 -- 2008 crisis period is shaded.}
\label{fig:covEU1}
\end{figure}

\begin{figure}[!h]
\centering
\includegraphics[width=\textwidth]{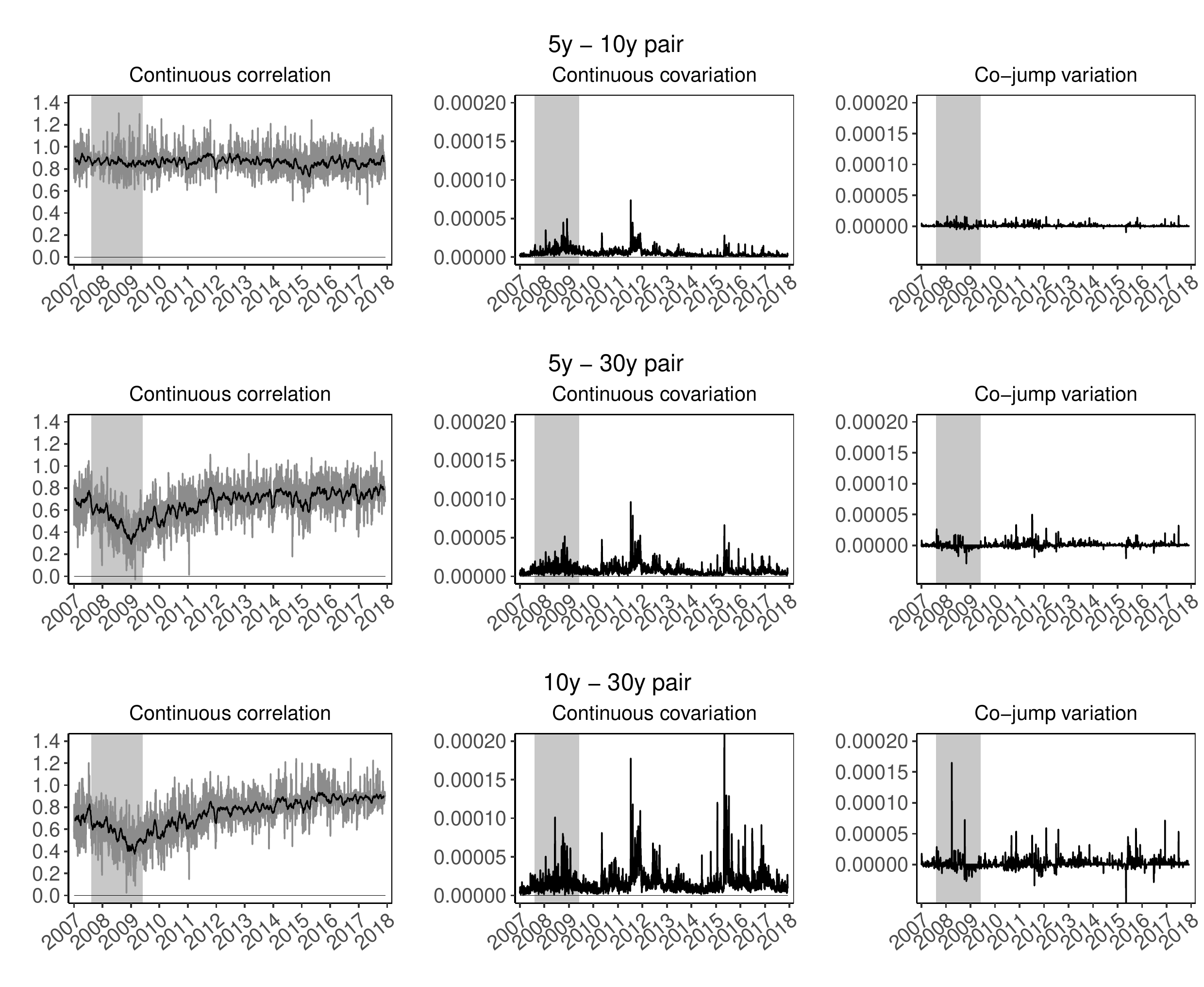}
\caption{European interest rates market: Decomposition of the covariation for the 5 year - 10 year, 5 year - 30 year, and 10 year - 30 year pairs. Continuous correlation in gray with a 21-day moving average in black (left column), integrated covariance (middle column), and co-jumps (right column) estimated by jump wavelet covariance estimator. The 2007 -- 2008 crisis period is shaded.}
\label{fig:covEU2}
\end{figure}

\clearpage
\pagebreak

\section{Mathematical Appendix}

\subsection{Discrete wavelet transform}
\label{dwt}

Here, we briefly introduce a discrete version of the wavelet transform. We use a special form of the discrete wavelet transform called the maximal overlap discrete wavelet transform (MODWT). We demonstrate the application of the discrete-type wavelet transform on a stochastic process using the pyramid algorithm \citep{Mallat98}. This method is based on filtering time series (or stochastic process) with MODWT wavelet filters and then filtering the output again to obtain other wavelet scales. Using the MODWT procedure, we obtain wavelet and scaling coefficients that decompose analyzed stochastic processes into frequency bands. Ror more details about discrete wavelet transforms and their applications, see \cite{PercivalMofjeld1997}, \cite{PercivalWalden2000}, and \cite{Gencay2002}.

The pyramid algorithm has several stages, and the number of stages
depends on the maximal level of decomposition $\mathcal{J}^m$. Let us
begin with the first stage. The wavelet coefficients at the first
scale $(j=1)$ are obtained via the circular filtering of time series $Y_{t,\ell}$ using the MODWT wavelet and scaling filters $h_{1,l}$ and $g_{1,l}$ \citep{PercivalWalden2000} :
\begin{equation}
\label{GrindEQ__13_}
\mathcal{W}_{1,t}^{\ell}\equiv \sum^{L{\rm -}{\rm 1}}_{l{\rm =0}}{}h_{1,l}Y_{{\rm (}t{\rm -}l\ modN{\rm )},\ell} \ \ \ \ \mathcal{V}_{1,t}^{\ell}\equiv \sum^{L{\rm -}{\rm 1}}_{l{\rm =0}}{}g_{1,l}Y_{{\rm (}t{\rm -}l\ modN{\rm )},\ell}.
\end{equation}
In the second step, the algorithm uses the scaling coefficients $\mathcal{V}_{1,t}^{\ell}$ instead of $Y_{t,\ell}$. The wavelet and scaling filters have a width $L_j=2^{j-1}\left(L-1\right)+1$. After filtering, we obtain the wavelet coefficients at scale $j=2$:
\begin{equation}
\label{GrindEQ__14_}
\mathcal{W}_{2,t}^{\ell}\equiv \sum^{L{\rm -}{\rm 1}}_{l{\rm =0}}{}h_{2,l}\mathcal{V}_{{\rm (1,}t{\rm -}l\ modN{\rm )}}^{\ell} \ \ \ \ \mathcal{V}_{2,t}^{\ell}\equiv \sum^{L{\rm -}{\rm 1}}_{l{\rm =0}}{}g_{2,l}\mathcal{V}_{{\rm (1,}t{\rm -}l\ modN{\rm )}}^{\ell}.
\end{equation}
The two steps of the algorithm create two vectors of the MODWT wavelet
coefficients at scales $j=1$ and $j=2$;
$\mathcal{W}_{1,t}^{\ell},\mathcal{W}_{2,t}^{\ell}$ and a vector of
the MODWT wavelet scaling coefficients at scale two
$\mathcal{V}_{2,t}^{\ell}$ that is subsequently used for further
decomposition. The vector $\mathcal{W}_{1,t}^{\ell}$ represents the
wavelet coefficients that reflect the activity at the frequency bands $f[1/4,1/2]$, $\mathcal{W}_{2,t}^{\ell}$: $f [1/8,1/4]$ and $\mathcal{V}_{2,t}^{\ell}$: $f[0,1/8]$.

The transfer function of the wavelet filter $h_l:l=0,1,\dots ,L-1$, where \textit{L} is the width of the filter, is denoted as $H(.)$. The pyramid algorithm exploits the fact that if we increase the width of the filter to $2^{j-1}\left(L-1\right)+1$, the filter with the impulse response sequence has the form:
\begin{equation}
\{h_0,\underbrace{0,\dots ,0}_{\mbox{$\scriptscriptstyle{2^{j-1}-1}$ \scriptsize{zeros}}},h_1,\underbrace{0,\dots ,0}_{\mbox{$\scriptscriptstyle{2^{j-1}-1}$ \scriptsize{zeros}}},h_{L-2},\underbrace{0,\dots ,0}_{\mbox{$\scriptscriptstyle{2^{j-1}-1}$ \scriptsize{zeros}}},h_L\},
\end{equation}
and a transfer function defined as $H\left(2^{j-1}f\right)$. Then, the
pyramid algorithm takes on the following form:
\begin{equation}
\mathcal{W}_{j,t}^{\ell}\equiv \sum^{L{\rm -}{\rm 1}}_{l{\rm =0}}{}h_l\mathcal{V}^{\ell}_{\left(j{\rm -}{\rm 1,}t-2^{j-1}l\ modN\right)}{\rm \ \ \ \ }t{\rm =0,1,\dots ,N-1,\ \ \ }
\end{equation}
\begin{equation}
\mathcal{V}_{j,t}^{\ell}\equiv \sum^{L{\rm -}{\rm 1}}_{l{\rm =0}}{}g_l\mathcal{V}^{\ell}_{\left(j{\rm -}{\rm 1,}t-2^{j-1}l\ modN\right)}{\rm \ \ \ \ }t{\rm =0,1,\dots ,N-1,\ }
\end{equation}
where in the first stage, we set $Y_t=\mathcal{V}_{0,t}^{\ell}$. After
applying the MODWT, we obtain $j\le \mathcal{J}^m\le log_{2}(N)$ vectors of wavelet coefficients and one vector of
scaling coefficients. The $j$-th level wavelet coefficients in vector
$\mathcal{W}_{j,t}^{\ell}$ represent the frequency bands $f [1/2^{j+1}{\rm ,1/}{{\rm 2}}^j{\rm ]}$, whereas the $j$-th level scaling
coefficients in vector $\mathcal{V}_{j,t}^{\ell}$ represent $f[0,1/2^{j+1}]$. In our analysis, we use the MODWT with the Daubechies
wavelet filter D(4) and reflecting boundary conditions.

\subsection{Bootstrapping the co-jumps}
\label{sec:bootstrap}

Under the null hypothesis of no jumps and co-jumps in the $(Y_{t,\ell_1},Y_{t,\ell_2})$ process,
\begin{eqnarray}
\mathcal{H}^0:\widehat{QV}^{(RC)}_{\ell_1,\ell_2}-\widehat{IC}^{(JWC)}_{\ell_1,\ell_2} &=& 0  \\
\mathcal{H}^A:\widehat{QV}^{(RC)}_{\ell_1,\ell_2}-\widehat{IC}^{(JWC)}_{\ell_1,\ell_2} &\ne& 0.
\end{eqnarray}
We propose a simple test statistic that can be used to detect significant co-jump variation. If a significant difference exists between the quadratic covariation and integrated covariance, then it is highly probable that we will observe a co-jump variation, possibly because of co-jump(s) or large disjoint jump(s). In this case, the $\mathcal{H}^0$ is rejected for its alternative.

When the null hypotheses of no jumps holds,
$\widehat{QV}^{(RC)}_{\ell_1,\ell_2}-\widehat{IC}^{(JWC)}_{\ell_1,\ell_2}$
is asymptotically independent from
$\widehat{QV}^{(RC)}_{\ell_1,\ell_2}$ conditional on the volatility path, and we can use two independent random variables to set the Hausman-type statistics to test for the presence of jumps. We proceed by scaling $\widehat{QV}^{(RC)}_{\ell_1,\ell_2}-\widehat{IC}^{(JWC)}_{\ell_1,\ell_2}$ by the difference in the variances of both estimators, which we obtain using a bootstrap procedure.

Under the null hypothesis of no jumps and co-jumps, we generate $i$ intraday returns $(r_{i,\ell_1}^*,r_{i,\ell_2}^*)$ with integrated covariance determined based on empirical estimates as
\begin{eqnarray}
r_{i,\ell_1}^*&=&\sqrt{\frac{1}{N} \widehat{IC}^{(JWC)}_{\ell_1,\ell_1}}\eta_{i,\ell_1} \\
r_{i,\ell_2}^*&=&\sqrt{\frac{1}{N}\widehat{IC}^{(JWC)}_{\ell_2,\ell_2}}\left( \widehat{\rho}_{\ell_1,\ell_2} \eta_{i,\ell_1}+\sqrt{1- \widehat{\rho}^2_{\ell_1,\ell_2}} \eta_{i,\ell_2}\right), \\
\end{eqnarray}
with $ \widehat{\rho}_{\ell_1,\ell_2}$ being the correlation obtained
from the $\widehat{\bm{IC}}^{(JWC)}$ matrix, and $\eta_{i,\ell_1} \sim \mathcal{N}(0,1)$ and $\eta_{i,\ell_2} \sim \mathcal{N}(0,1)$.
Now, we use $(r_{i,\ell_1}^*,r_{i,\ell_2}^*)$ to compute $\widehat{QV}^{(RC)*}_{\ell_1,\ell_2}$ and $\widehat{IC}^{(JWC)*}_{\ell_1,\ell_2}$. Generating $b=1,\ldots,B$ realizations, we obtain $\mathcal{Z}^*=(\mathcal{Z}^{(1)},\mathcal{Z}^{(2)},\ldots,\mathcal{Z}^{(B)})$ as
\begin{equation}
\mathcal{Z}^*=\frac{\widehat{QV}^{(RC)*}_{\ell_1,\ell_2} - \widehat{IC}^{(JWC)*}_{\ell_1,\ell_2}}{\widehat{QV}^{(RC)*}_{\ell_1,\ell_2}}.
\end{equation}
which can be used to construct a bootstrap statistic to test the null hypothesis of no co-jumps as
\begin{equation}
\mathcal{Z}=\frac{\frac{\widehat{QV}^{(RC)}_{\ell_1,\ell_2} - \widehat{IC}^{(JWC)}_{\ell_1,\ell_2}}{\widehat{QV}^{(RC)}_{\ell_1,\ell_2}}-E(\mathcal{Z}^*)}{\sqrt{Var(\mathcal{Z}^*)}}\sim\mathcal{N}(0,1).
\end{equation}
The bootstrap expectation and variance depend on the data. We rely on
the assumptions of \cite{dovonon2014bootstrapping}. Thus, by identifying days when the co-jump component is present, we can estimate the off-diagonal elements of the covariance matrix $\widehat{\bm{IC}}^{(JWC*)}$ as
\begin{equation}
\label{iccxy}
\widehat{IC}_{\ell_1,\ell_2}^{(JWC*)} = \widehat{QV}^{(RC)}_{\ell_1,\ell_2}\mathbbm{1}_{\{ \vert\mathcal{Z}\vert \le \phi_{1-\alpha/2}\}}+\widehat{IC}^{(JWC)}_{\ell_1,\ell_2}\mathbbm{1}_{\{ \vert\mathcal{Z}\vert > \phi_{1-\alpha/2}\}},
\end{equation}
where $\phi_{1-\alpha/2}$ is a critical value for the two-sided test
with a significance level $\alpha$. Finally, we estimate all elements of the (continuous) covariance matrix:
\begin{equation}
\widehat{\bm{IC}}^{(JWC*)}=\left(
\begin{array}{cc}
\widehat{IC}^{(JWC*)}_{\ell_1,\ell_1} & \widehat{IC}_{\ell_1,\ell_2}^{(JWC*)}\\
\widehat{IC}_{\ell_2,\ell_1}^{(JWC*)} & \widehat{IC}^{(JWC*)}_{\ell_2,\ell_2}
\end{array}
\right).
\end{equation}

\end{document}